# Near-Peer Mentoring in Data Science: A Plot for Mutual Growth

Chiara Sabatti
Departments of Biomedical Data Science and of Statistics,
Stanford University, Stanford, CA
and
Qian Zhao
Department of Mathematics and Statistics,
University of Massachusetts Amherst, Amherst, MA

August 15, 2025

**Abstract**

Universities have been expanding undergraduate data science programs. Involving graduate students in these new opportunities can foster their growth as data science educators. We describe two programs that employ a near-peer mentoring structure, in which graduate students mentor undergraduates, to (1) strengthen their teaching and mentoring skills and (2) provide research and learning experiences for undergraduates from diverse backgrounds. In the Data Science for Social Good program, undergraduate participants work in teams to tackle a data science project with social impact. Graduate mentors guide project work and provide just-in-time teaching and feedback. The Stanford Mentoring in Data Science course offers training in effective and inclusive mentorship strategies. In an experiential learning framework, enrolled graduate students are paired with undergraduate students from non-R1 schools, whom they mentor through weekly one-on-one remote meetings. In end-of-program surveys, mentors reported growth through both programs. Drawing from these experiences, we developed a self-paced mentor training guide, which engages teaching, mentoring and project management abilities. These initiatives and the shared materials can serve as prototypes of future programs that cultivate mutual growth of both undergraduate and graduate students in a high-touch, inclusive, and encouraging environment.

# 1 Introduction

The education literature documents multiple advantages of training opportunities where graduate students work with undergraduates in mentoring capacities. First, involving graduate students can help decrease the student-to-instructor ratio in many courses (Sorte et al., 2020). Second, the access to mentors increases retention of first generation and under represented minorities (URM) in Science, Technology, Engineering and Medicine (STEM) disciplines (Wilson et al., 2012). Graduate students are *near-peer mentors*: while more advanced in training and education, they are of a similar age to their mentees (Garcia et al., 2021). Near-peer mentoring in particular has been shown to be an effective mechanism with which to foster a sense of belonging and self-identity as a scientist (Zaniewski and Reinholz 2016; Wilson and Grigorian 2019). Third, acting as mentors positively influences the personal, cognitive, and professional growth of graduates student mentors (Dolan and Johnson 2009; Tenenbaum et al. 2014). Opportunities that value interdependence and recognize communal goals may increase some students' cultural alignment with the research institution (Stephens et al. 2012; Boucher et al. 2017; Anderson et al. 2019; Belanger et al. 2020).

The possible benefits of near-peer mentoring resonate particularly well with two challenges we encounter in data science education: the need to increase access to research experience, and to diversify the workforce. First, while both the Guidelines for Assessment and Instruction in Statistics Education (GAISE) Reports (2016) and Curriculum Guidelines for Undergraduate Programs in Statistical Science (2014) recommend research experiences as crucial means of enhancing learning, limited faculty time to mentor can impede the development and expansion of such opportunities (Nolan et al., 2020). Engaging graduate students in mentoring roles can increase universities' capacity to offer these project-based and experience-centered learning (Rodolfa et al., 2019).

Second, while it is well recognized that increasing diversity expands the talent pool, boosts innovation and improves the nation's global leadership in science and technology (National Academy of Sciences, Engineering, and Medicine, 2011), students who identify as female or belonging to minority groups remain a lower proportion of doctorates in the field of statistics (National Center for Science and Engineering Statistics, 2022). Since mentor-ing experience increases the sense of belonging, self-efficacy, and science identity for both underrepresented mentors (Trujillo et al., 2015) and mentees (Zaniewski and Reinholz, 2016; Wilson and Grigorian, 2019), it can potentially improve retention at both undergraduate and graduate levels. The strong job market in data science Davenport and Patil (2012), which is projected to grow 36% in the decade between 2023-2033 (U.S. Bureau of Labor Statistics, 2025), offers concrete opportunities for the students, possibly from URM, towards which we direct outreach and retention efforts.

While research, consulting, or capstones are offered in many departments and they have been described in publications (e.g., Lazar et al. (2011); Smucker and Bailer (2015); Legler et al. (2010)), literature on statistics programs that engage graduate students in the role of mentors is lacking. In this article, we report on two programs developed at a four-year university, relying on a near-peer mentoring structure to provide strong learning experience for both undergraduate and graduate students. In the Data Science for Social Good (DSSG) summer fellowship program, graduate students mentor undergraduate student teams to tackle a data science project with positive social impact over the course of eight weeks. Stanford Mentoring in Data Science (SMDS) is both a class offered to graduate students and an outreach initiative. Mentors enrolled in this class participate in interactive workshops focused on inclusive mentoring as well as engage in one-on-one meeting with undergraduate mentees. Both programs were introduced in response to graduate students' interests, and graduate students continue to actively take part in program design. Faculty instructors developed program goals and structure; they support the graduate mentors, creating a robust learning experience for both undergraduate and graduate participants. We describe our experiences with the two programs (Sections 2 and 3), and share learning materials we have developed for graduate student mentors (Section 4). We hope that this will encourage and support future near-peer mentoring programs, fostering learning and a sense of belonging for both graduate and undergraduate students.

## 2 Data Science for Social Good

Data Science for Social Good (DSSG) is summer fellowship where student teams work on data science projects with social good objectives. The impetus of DSSG came from a request by a graduate student who participated in the homonymous program at the University of Chicago and, having found the experience transformative, was determined to create some version of it at our university. Motivated by their testimony and willingness to work towards this goal, we ran a pilot experience and then develop the program over four years. Compared to other DSSG programs nationwide (see Dailey et al. (2021), for a review), our program is unique in that its core objective is to provide educational opportunities for both undergraduate participants and graduate mentors in distinct roles.

## 2.1 Program goals

The DSSG program serves three groups: undergraduate participants, graduate student mentors in training, and project partners.

Our undergraduate participants are "beginner data scientists," who have sufficient technical competencies, that is, have taken upper undergraduate courses in statistical inference, linear regression, and have experience using statistical software through at least one course, but without experience applying data science to a real world problem, and may not be committed to a career in this space. We set four broad goals for this group: (1) to engage in an authentic data science project that addresses a societal need; (2) to receive training, mentoring, and practical experience in aspects aspects of "data acumen" identified by National Academy of Sciences, Engineering, and Medicine (2018) least reflected in a typical educational curriculum, e.g., ethics, communication, and collaboration; (3) to experience the feeling that data scientists are "people like me"; (4) to strengthen professional skills and increase self-efficacy, feeling empowered to make career choices with a sense of purpose and service. We focus on the following questions when selecting participants: Do they have adequate foundations in data science, statistics and programming? Are they excited to apply data science for social good? What research areas and society issues are they passionate about? What are their experiences with working in a diverse team? What are their commitment to diversity and inclusion?

Our mentors in training are graduate students and postdocs, who have strong technical expertise and some training in teaching (for example, serving as graduate teaching assistants (TA), or through TA workshops), but limited experience of mentoring, managing a research team, or navigating the challenges of interdisciplinary work. By engaging in DSSG, the mentors build competence in the following areas: (1) mentoring undergraduate participants in an “active learning” environment; (2) lead a data science team project; (3) engage with project partners of different background without the scaffolding provided from an instructor. We select mentors through an open application, encouraging responses in particular from students supported by the data science initiative, and enrolled in PhD programs as Statistics, Computer Science, Biomedical Data Science.

Our project partners are often non-profit organizations who have a research question that can be addressed through data analysis. Faculty from across the university is invited to contribute project ideas and suggest potential partners. Potential partners's expectations regarding scope and impact are framed with the understanding that undergraduate students carry out the majority of the work.

## 2.2 Roles of graduate mentors

At DSSG, graduate mentors in training take the lead in project development and participants' learning. We now outline in some detail the activities they engage in, clarifying how these strengthen their teaching, mentoring, project management, and leadership skills.

**Identify project goals and milestones** Mentors and faculty advisors meet with the project partner prior to the program to achieve a broad vision of the needs and potential of the collaboration. Mentors then scope the project, keeping in mind its societal impact, the learning it provides to participants, feasibility given the time frame, and the ease with which it can be communicated. Working with the project partner, mentors define concrete goals, weekly milestones, and clarify the available resources. An example project milestones is provided in the Supplementary Material. Mentors typically do not curate data beforehand, in order to expose participants to real-world data and encourage them to take a leading role in determining which aspect of the data can be used to address which part of the challenge.

**Project onboarding** Mentors develop an onboarding document, which includes background information, related literature, and technical resources. At the start of the program, participants go through the material, meet with the project partner, and ask clarifying questions. After the meeting, participants frame their overarching goal for the summer and identify their first set of tasks, which often involve identifying and processing relevant data and exploratory data analysis.

**Technical training** Mentors provide technical training to all participants, such as how to write a reproducible document in R or how to use cluster computing, as well as topics that arises during project work. An example of training materials that mentors developed to teach SQL database is provided in the supplement. Mentors also facilitate participants' self-directed learning, guiding them to suitable resources. Through these experiences, mentors learn how to teach in an active learning environment, encourage explorations while integrating learning into project work, support a growth mindset, and foster students' confidence.

**Facilitate team meetings** During the program, mentors participate to the daily project team meeting. Discussion topics range from brainstorming potential methods to evaluating different approaches, from presenting to a non-technical audience to developing reproducible workflow. Mentors can also use daily team meeting to provide just-in-time teaching (McGee et al., 2016; Simkins and Maier, 2010), e.g., correct misconceptions, or pose questions for discussions. This regular interaction provides an opportunity for the mentor to facilitate successful team work, coordinate a team project, and foster a sense of belonging.

**Collaborate with project partners** Each week, the project team meets with faculty advisors and project partner to share progress, collect feedback, clarify directions, and identify areas where the DSSG team needs support. Both the mentor and participants learn to collaborate with domain experts who do not have strong data science background, but can articulate relevance and implications of the project.

**Build a data science community** Mentors organize activities to introduce participants to the wide range of data science topics and career paths. For example, members of the data science community are invited to describe their career trajectory or projects in a weekly presentation. Mentors also organize social activities such as improvisation workshop and trivia challenge to cultivate a sense of community among participants. An example schedule of the DSSG program is included in the Supplementary Material. Mentors meet weekly with the faculty advisors to discuss mentoring challenges and make plans for program activities.

## 2.3 Program outcomes

We have run the DSSG program for seven summers, between 2019 and 2025. Data in this report refers to the first five. The average number of people involved has been 8 participants, 3 mentors and two faculty members. Demographics details on participants and mentors from 2019–2022 are available in Figure 1 in the Supplementary Material and the partner and project details can be gathered from the program web-pages. We achieved our goal of involving underrepresented minorities: about 50% of participants and 70% of mentors identify as female or other gender identity. More than 70% of participants and 60% of mentors are from a minority group as described in Figure 1.

We have also surveyed DSSG mentors on their experience. Mentors who responded to the anonymous online survey ($n = 9$) reported gain in their capacity in multiple dimensions (see Appendix A). For instance, a vast majority are more confident in their ability to employ different teaching strategies, working with students to set academic and professional goals, setting goals and milestones of a data science project, and communicating progress with other participants and project partners. To quote one of the mentors: "Before DSSG, although I have served as a TA for several classes, I have never had experience mentoring junior students. I learned from my experience that different people have different ways to approach projects or respond well to different mentoring styles, which can be vastly different from mine. I also learned a bit about how to structure team work such that students with different strengths can complement or learn from each other."

This comment highlights that the DSSG can be a significant learning experience for mentors. It also shows that mentoring a team project can be an unfamiliar and at times challenging for graduate students. As we continue developing this program and other courses that are centered around undergraduate research projects, we start to recognize the importance to meaningfully include and engage graduate students who may not have prior experience or consider themselves as ideal mentors. With the goal of supporting graduate mentors in mind, we have developed a set of learning materials for DSSG mentors, that can be applicable to mentors of general data science projects. We describe the guide in detail in Section 4 and discuss the second program in the next section.

# 3 Stanford Mentoring in Data Science

Stanford Mentoring in Data Science (SMDS) doubles as a class offered to graduate students and an outreach initiative. Graduate students enrolled in the class participate in both weekly in-class meetings and a mentorship practicum. SMDS was launched during academic year 2020-21, when instruction was entirely remote, and the US society had experienced a number of dramatic events. In particular, the protests and reckoning with racism had engaged the graduate students population, who self-organized in many forms of activism. A program aimed at increasing the diversity of the data science workforce aligned well with their interests and offered them an opportunity to channel their desire for action. We describe the program goals for both graduate students and undergraduate participants in Section 3.1 and outline program structure in Section 3.2.

## 3.1 Program goals

The ultimate goal of SMDS is to assure that the data science workforce is well connected to society and reflects its diversity. The program works towards this goal with a dual strategy. On the one hand, SMDS recruits undergraduate students from currently underrepresented groups in data science, connecting them to the field. On the other hand, SMDS educates graduate students on values of diversity in a research environment, equipping them with techniques and practice to mentor individuals from diverse backgrounds.

The undergraduate participants are at the early stages of their exposure to data science. For example, they may have taken an introductory statistics course, or they might be more advanced in a related program but lack hands-on experience. We aim to motivate, support and nurture their interest in data science through mentoring and introduction of career paths and research topics. To reach out to potential mentees, we have contacted community colleges from Carnegie Math Pathways: participating faculty advertises the program to their students and directly refer candidates. The application form is purposely simple, collecting information on demographics, current school, courses completed in math/computer science/statistics, areas of interest, and plans for the future. When selecting participants, we look at questions like: Do they have enough background to engage in some data exploration? Is there room in their plans to increase and sustain an interest in data science? Do they appear to be in need of mentorship?

Graduate students working in the data science space can enroll into a class (BIODS360), offered for credit. They learn (1) to develop a set of tailored goals for the mentoring relationship based on the needs and aspirations of their mentees; (2) to recognize stereotypes and implicit biases that might shape our behavior and learning and to utilize strategies that overcome these, creating and maintaining an inclusive environment; (3) to articulate the value of a diverse and inclusive team in tackling data science problems, being informed about current composition of data science workforce, with an eye both on historical context and future goals for our field.

## 3.2 Program structure

For the duration of the quarter (ten weeks), graduate students enrolled in the course participate weekly to one in-person class and meet one-on-one online with their undergraduate mentees. In-class meetings consist of interactive workshops, reflection on mentoring experience, and community building activities.

During the first in-class meeting, after an introduction of the course structure, graduate students are divided in pairs and go through "the wallet exercise." Designed by the Stanford d.school (see references for links to instruction and facilitator's guide), this workshop task participants to design an "ideal wallet" for their partner. After interviewing their partner, students brainstorm an initial design, which they refine after iterations of feedback, and eventually come up with a prototype. This activity serves multiple purposes: it acts as an icebreaker, helping to establish a climate for safe exploration and sharing; it underscores the importance of listening and empathy during meetings with participants; it frees mentors from the idea of an "ideal" project and encourages them to jointly develop one experience that fulfilled their mentee's needs; it introduces the idea that false starts are possible and sometimes necessary to find the right path.

The first class concludes with the assignment for each graduate student to identify their mentee, respond to the e-mail that the course organizer would send to the pair establishing contact, scheduling a first meeting during the current week. During the first meeting graduate students are tasked to learn as much as possible about their mentee, share some information about themselves to start developing a relationship, and find a mutually agreeable time to meet online each week. We found that letting graduate mentors self-match to undergraduate participants, based on some of the information collected in the application, e.g., name, gender, degree, and area of interest, increased mentors' sense of agency without creating specific biases.

During the second in-class meeting, all participants are invited to join in on Zoom. Everyone briefly introduced themselves, with the aid of icebreaker questions (e.g., "if I were an animal I would be a..., because ..."). The instructors of the class then share examples of directions that the mentee and mentors could take in their work together: preparing for internship applications, learning a data science topic, exploring careers, taking a close look at a dataset on a topic of interest etc.. While we try to underscore that each pair had total autonomy in deciding their "curriculum," we have found that being presented with a list of possibilities helped them to articulate needs and interests. The assignment for the second one-on-one meeting is for the mentor and participant to do a first round of brainstorming on activities that they might engage in during the program. We suggest keeping this brief, leaving time to re-evaluate during the coming weeks. During the rest of the meeting, we invite mentors to share a bit more on their experience as data scientists, describing the course of study they are involved in and the topic of their dissertation.

In the following weeks, we invite a number of campus experts in the area of diversity, equity and inclusion to facilitate workshops. They introduce and discusse concepts as belonging, identity interference, stereotype threat, imposter syndrome, offering pointers to literature (Leslie et al. 2015; Stephens et al. 2012) as well as personal experiences. A list of workshops we have offered are included in the Supplementary Material. We found these workshops particularly useful for both the program and graduate participants. Engaging with experts provided a strong signal of the importance of these topics and the amount of scholarly work that has been carried out. Speakers brought to the table years of personal experience which grounded the concepts and solutions presented and established warmth and acceptance in discussions. Connecting with campus resources substantially enlarged the reach of the program, as we continued to leverage these in other contexts.

A number of in-class meetings are devoted to mutual sharing of how the mentor-mentee relationships are evolving, and brainstorming around possible roadblocks. The instructors meet one-on-one with undergraduate participants and bring to the meeting common feedback from participants. By hearing from each other, mentors are better able to notice their own growth and identify successes and challenges. They are more motivated to develop solutions knowing that they will benefit others. Sharing and discussing as a group also fosters a sense of community among graduate students from different academic backgrounds.

Towards the end of the program, we invite all participants to join a synchronous class meeting. This is an opportunity for participants to share their growth during this program, reinforce peer connections, and present ways to continue the peer-relationships. We have organized career panels during one of the classes, inviting data science professionals who had attended the same schools as our participants. Due to the pandemic and available resources, our participants' experience has been entirely online so far, and we are working to organize an in-person gathering for the conclusion of the program.

Graduate mentors have created a collection of resources over the years: pointers to outreach programs and internships; articles on data science careers; suggestions on resume and cover letters; tutorials on statistics topics and software; and a collection of datasets and basic analysis scripts in R and Python that can be used as a starting point of data exploration and analysis. An example project can be found in the Supplementary Material. These resources help to take the pressure off mentors to search for or create resources, allowing them to focus on mentoring.

## 3.3 Program outcomes

We have offered SMDS five times during 2021–2025. A summary of the demographic and education information for participants and mentors in the first three offerings is provided in Figure 2 in the Supplementary Material. Nearly half of the undergraduate participants are from underrepresented minority groups in science and engineering professions. We surveyed participants' program experience and attitude in end-of-program anonymous online questionnaires. During meetings with their mentor, mentees often spent time on learning data science skills (59% of time on average), career counseling (30% of time on average), and individual projects (30% of time on average). End-of-program surveys show mentees became more interested in data science and in pursuing future internships. Graduate students' feedback is collected via the University official course evaluation system. The program has been reported to increase mentors' interest to engage in activities to enhance diversity in their working environment. We report survey results in Appendix B.

# 4 A guide for Data Science for Social Good mentors

Research have shown that mentor training improves mentor's skills and self-efficacy (Stamp et al., 2015). Moreover, it is known that socioemotional and culturally relevant mentoring positively impacts mentee's research experiences (Haeger and Fresquez, 2016). These findings, together with our intention to include and engage mentors from diverse academic and personal backgrounds, motivated us to develop a self-paced learning guide, *Growing by mentoring: A guide for Data Science of Social Good mentors*, to share our learning in the two programs and provide support to DSSG project mentors. While grounded in DSSG projects, the themes and activities are broadly applicable to graduate student mentors of data science project teams.

*Growing by mentoring: A guide for Data Science of Social Good mentors* covers three broad areas: (1) what are characteristics of a good mentor, and how to develop them; (2) how to facilitate the progress of a team working on an interdisciplinary data science project; (3) how to set individualized learning goals and approaches to help students reach them. Organized around these themes, the guide introduces relevant concepts and useful strategies. Each idea is made concrete with an "activity": by completing all of these, mentors will be able to carry out most of the tasks they need to design their mentoring experience. Each section is self-contained, such that mentors can pick and choose specific topics of interest. The guide is included in the Supplementary Material, and we outline main topics and activities in this section. While we are aware of established research mentor training programs (Pfund et al., 2014), this is perhaps the first set of self-paced materials developed specifically for graduate mentors for data science projects.

## 4.1 Growth as a mentor

The first section prepares graduate students to become *intentional mentors*, who have visions of for both themselves and their mentees and plan their course of actions accordingly. This section comprises four topics: goal setting, effective mentoring, belonging, and creating an inclusive learning environment.

We introduce mentoring as a type of support intended to advance mentees toward an important goal (Packard, 2016). It is a reciprocal relationship where both the mentee and the mentor can experience personal, professional, and intellectual growth from respectful, effective, and sustained interactions (Galbraith, 2003). We invite readers to consider the qualities and actions of their own mentors. This reflection serves as an entry point into discussions of attributes and practices of effective mentors. We leverage the triangular model of mentor competence from the book *On being a mentor* (Johnson, 2007) and examples from the article *Nature's guide to mentors* (Lee et al., 2007). To conclude, we invite mentors to reflect on their own strength and weaknesses, and set their mentoring goals. This aims to cultivate a growth mindset, encouraging a continued development of these skills throughout the program.

The objective of mentoring is to support mentees to make progress towards their goals. We describe the S.M.A.R.T. (Specific, Measurable, Achievable, Relevant, and Time-Bound) goal setting framework, provide a list of data science skills students may want to improve in, adapted from *The data for good growth map* (Dailey et al., 2021) and Curriculum Guidelines for Undergraduate Programs in Statistical Science from American Statistical Association Undergraduate Guidelines Workshop (2014), and a list of potential questions, to clarify the breadth of topics mentors can discuss with their mentees.

The next two sections raise mentors' awareness of challenges minority students can face and strategies they can use to create an inclusive learning environment. We introduce the concept of belonging from Walton and Wilson (2018) and highlight the importance of belonging in a learning environment. We discuss challenges to belonging for under-represented students, such as identity interference (Slepian and Jacoby-Senghor, 2021), stereotype threat (Steele and Aronson, 1995) and previous negative experiences. We then offer strategies to enhance belonging, such as through reframing challenges (Walton and Wilson, 2018) and cultivating a growth mindset (Dweck, 2006). We refer to resources such as Lang (2016) and Arif et al. (2021). Mentors watch online videos where students speak about issues they have faced, and brainstorm how they can implement these strategies in their projects.

## 4.2 Growth of the project

Group projects can offer a rich learning experience for undergraduates to analyze complex data, think in context, and choose appropriate method as advised in Curriculum Guidelines for Undergraduate Programs in Statistical Science (2014). Leading a collaborative data science project can strengthen mentors' leadership and project management skills. This section intends to prepare mentors in four aspects: initiating and sustaining collaboration, setting concrete project milestones, facilitating discussions, and organizing teamwork.

We borrow from research in meaningful collaboration in community-based participatory research (CBPR). In both CBPR and Data Science for Social Good projects, the objective is often to spawn positive changes, the project is often a systematic inquiry, and both the data science team and the project partner are engaged throughout the process. Mentors first reflect on a case study, to help them contextualize the discussion and set realistic expectations. We then offer a list of suggestions, based on Pasick et al. (2010) and Swann and Murray (2020), accompanied by an example of a DSSG or CBPR project. For instance, we suggest mentors to be flexible and open to alternative approaches. As an example, we describe a DSSG project where the team needed to develop a specialized workflow to use highly sensitive data, and they embraced the challenge as an opportunity to learn privacy consideration and writing efficient code (Foster et al., 2021).

In another activity, mentors are invited to identify overarching objectives and create project milestones. We provide questions they can discuss with their project partner, emphasizing the importance to clarify the need to be addressed, to analyze the impact on communities who may be affected, and to draft milestones. We borrow ideas from the book *Ten steps to successful project management* (Russell, 2007).

Unlike project managers who assigns tasks to team members, mentor's role is to facilitate students' learning and discovery. Team discussion serves as a medium where students collectively find a way forward, generate and express insights and discoveries. The third section encourages mentors to be more deliberate and purposeful when facilitating discussions. We present two perspectives of discussion: components (questioning, listening and responding, following up), and structures (framing, encouraging deeper answers, broadening perspectives, refocusing, and closing the loop). These materials are adapted from workshops and resources from centers for teaching and learning from multiple institutions.

Finally, we encourage mentors to form a consensus with the team about project organization and team collaboration, e.g., how to keep track of each other's progress, and how to organize code. Together, the four sections serve as a starting point as mentors embark on their journey as a project leader and manager.

## 4.3 Growth as a teacher

The third section aims to support mentor's growth as a data science educator by providing tools and practices to articulate learning goals, cultivate an active learning experience, and assess students' learning. This section is largely based on two books: *How learning works: Seven research-based principles for smart teaching* (Ambrose et al., 2010) and *The ABCs of how we learn: 26 scientifically proven approaches, how they work, and when to use them* (Schwartz et al., 2016).

We introduce both the cognitive aspects of learning using the revised Bloom taxonomy (Anderson et al., 2001) and the more holistic perspective of learning using Fink's taxonomy (Fink, 2013). Mentors are then invited to refine learning goals they have set for mentees from these perspectives.

Next, we focus on three approaches mentor can use to faciliate learning during project work: (1) cultivating a collaborative learning environment, (2) fostering meta-cognitive skills, and (3) incorporating active learning. Again, the introduction of relevant concepts is accompanied by concrete data science examples, and activities the mentors can leverage in their program. For instance, we introduce meta-cognition as the process of reflecting on and directing one's own thinking (National Research Council, 2001). Then, we describe the steps of problem solving as well as questions mentors can ask students to reflect on each of these steps, following the characterizations by Price et al. (2021). Mentors are invited to come up with additional questions and ask them during team discussions. By modeling the metacognitive process and gently nudging students to reflect on their learning and problem solving process, mentors can help students develop habits with long lasting impact. Taken together, this section serves the dual purpose of enhance mentor's skill as an educator and students' skill as a data scientist in training.

# 5 Discussion

In exit surveys, participants and mentors alike report an overwhelming positive experience in DSSG or SMDS. While we have collected feedback from participants and mentors, we have not run a case-control study, and analysis of the outcomes is additionally complicated by the fact that we are here reporting results of pilot programs, whose structure has evolved over time, in response to the feedback received. In addition, our main goals of long-term involvement in data science and education is impossible to measure in this short time frame. As these programs take a more permanent place in the offerings in our university, it will be possible to engage in more systematic outcome measurements in the future.

By sharing our experience and developed materials, we aim to facilitate the design of related programs that we imagine centered on the unique context and existing infrastructure in each institution. For universities with existing consulting or interdisciplinary study programs, giving some advising and mentoring role to graduate students, with faculties supervisors meeting weekly with undergraduate and graduate students can be a way to start applying a near-peer mentoring structure. For colleges without a graduate student population, upper-undergraduate students who have more research experience might serve the role of near-peer mentors, with additional structured guidance by faculty advisors.

We conclude this article with some personal reflections on the impact of these program on our experiences as a faculty member and as a graduate student. By design, DSSG and SMDS had graduate students and faculty members working side-dy-side in the organization and execution of programs in the educational space. This was an energizing and enriching experience, as both groups engaged outside the traditional roles of instructor and teaching assistant. Graduate students contributed energy and fearlessness in pursuing projects and opportunities, and were able to articulate and channel the expectations of the participants.

Faculty made sure that there was a fertile ground to sustain this growth: they were able to estimate and garner the resources needed, provide a referential framework with which to evaluate efforts and progress, place the activities in a broader educational context, bring in experts and pointing to methods and research areas. This enhanced collaboration improved the cohesion between faculty and students in the same department, and increased the agency and confidence of graduate students.

Working in a university, and having only a finite amount of time and resources, we experience occasionally tensions between what appear to be different goals: educate a new generation, transmitting the knowledge we have and cultivating in them the aspiration to keep increasing it; engage in research, pushing the boundaries of our understanding; deploy this knowledge in practical applications, impacting the society around us. Being involved in these programs was so enjoyable partly because multiple of these goals aligned: we had the opportunity to connect with many different reality on campus, learn, educate, and make an impact—university at its best.

## Acknowledgement

The value of experiences we described stems from the people who engaged in them: we are thankful to all participants, mentors, faculty and staff who dedicated time and energy to pilot and develop these programs. We thank Ben Stenhoug for his vision to create a Data Science for Social Good program at Stanford and Ben, Emily Flynn and Michael Sklar for agreeing to work as mentors in the first edition, very much pulling ourselves up from our bootstraps, and making a blue print for years to come. Thanks to Balasubramanian Narasihman for having shouldered major responsibilities for this program for seven years, committing all his summers and his good spirits, and being ready to join into the SMDS adventure in unflappable manner. Saara Khan and Jonatan Feiber, from the d.School, have enriched our experience with workshops and masterful handling of interactions over Zoom. Lucy Bernholz provided a much needed connection with the humanities and social science camp on campus. Connor Doherty coordinated many students contribution to SMDS. Miranda Stratton, Marcella Anthony, Joseph Brown, Zandra Jordan volunteered their time and expertises to help us grow our inclusivity muscles. Elizabeth Munoz, Porn-prang Plangsrisakul and Natalie LaMariana kept us, and a lot of people outside our university, on track. Bernadine Chuck Fong connected us with an incredible number of partners, starting from the Carnegy Math Pathways, where Dan Ray has been very helpful. Support from the National Science Foundation (NSF 1934578, DMS-2210392) and Stanford Data Science made all of this possible.

## Conflict of interest statement

The authors report there are no competing interests to declare.

# A DSSG mentor growth

We surveyed DSSG mentors who participated between the years 2019–2022. Each question is framed as following: "How does DSSG change your confidence in the following aspects of teaching data science (mentoring/project management/leadership and community building)?" Mentors are asked to rate on a Likert scale ("Much less confidence", "Somewhat less confident", "No change", "Much more confident", "Somewhat more confident") with an additional option of "Prefer not to respond". We code the five options using numeric values $(-2, -1, 0, 1, 2)$. Nine mentors took the survey, and almost all of them responded to all of the questions in the range from 0 to 2 (response rate = 70%). We report the mean and standard deviation of the responses to each aspect in Table 1. We note that the survey was first conducted in June, 2022, which may be a long time away from when some of the mentors participated in DSSG.

# B SMDS program feedback

We surveyed both participants and mentors who participated in SMDS at the end of the program. Altogether, 34 out of 57 undergraduate participants in 2021–2023 have responded to the survey (response rate = 60%). We received 26 responses from the graduate student mentors (response rate = 70%). Because the survey is anonymous, if a mentor participated in the program multiple times, their responses in each year are counted as separate responses. Participants and mentors are asked to rate on a Likert scale ("Definitely less interested", "Slightly less interested", "No change", "Slightly more interested", "Definitely more interested") and we code these five options using numeric values (-2, -1, 0, 1, 2). All of participants and mentors rated all of the questions in the range from 0 to 2. We report the mean and standard deviation of the responses to each aspect in Table 2.

Table 1: DSSG mentors' self-identified growth. Mentors rated their change of confidence in each category on a Likert scale, where -2 indicates "much less confident", 0 indicates "No change", and +2 indicates "Much more confident". We report the average of mentors' responses and its standard deviation within the parenthesis.

| Topics | Specific aspects | Mean |
|---|---|---|
| Teaching Data Science | T1. Assessing my mentee's prior knowledge and skills in data science. | 1.1 (0.2) |
| | T2. Communicating data science topics effectively. | 1.2 (0.3) |
| | T3. Employing different teaching strategies (e.g., worked example, peer feedback). | 1.3 (0.2) |
| | T4. Encouraging students to reflect on their work. | 0.9 (0.3) |
| Mentoring | M1. Connecting with students from diverse backgrounds. | 1.1 (0.3) |
| | M2. Cultivating a growth mindset. | 0.9 (0.2) |
| | M3. Working with mentees to set academic and professional goals. | 1.3 (0.2) |
| | M4. Providing constructive feedback. | 1.0 (0.2) |
| | M5. Actively listening to mentees' feedback and concerns, and incorporating them into my mentoring. | 1.3 (0.3) |
| Project Management | P1. Gauging the impact of a project. | 1.2 (0.1) |
| | P2. Setting goals and milestones for a data science project. | 1.4 (0.2) |
| | P3. Communicating progress with participants and community partners. | 1.1 (0.3) |
| Leadership, Community building | L1. Facilitating group collaboration. | 1.2 (0.1) |
| | L2. Motivating and engaging participants. | 1.0 (0.2) |
| | L3. Building a community of data scientists. | 0.8 (0.3) |

Table 2: Participants and mentors' feedback at the end of SMDS program. Participants and mentors rated their change of interest in each question on a Likert scale, where -2 indicates "Definitely less interested", 0 indicates "No change", and +2 indicates "Definitely more interested". We report the average of responses and its standard deviation within the parenthesis.

| | Questions | Average |
|---|---|---|
| Participants | T1. What portion of the time was spent on on learning specific data science skills? | 59% (4%) |
| | T2. What portion of the time was spent on career counseling (e.g., CV revision, interview preparation, class selection, etc.?) | 30% (5%) |
| | T3. What portion of the time was spent working on a project? | 29% (4%) |
| | P1. How has participating in this experience changed your interest in Data Science? | 1.6 (0.1) |
| | P2. How has your experience in this program changed how likely you are to apply for other internships/mentorships? | 1.7 (0.1) |
| Mentors | M1. How has participating in this class changed your interest in mentoring? | 1.4 (0.1) |
| | M2. How has participating in this class changed your willingness to engage in activities aimed at increasing diversity in the groups of people with whom you work? | 1.5 (0.1) |

# SUPPLEMENTARY MATERIAL

**DSSG program materials** Descriptions of activities; example project milestone; example onboarding material; example of learning material.

**SMDS program materials** List of SMDS workshop topics; example of analysis scripts.

**DSSG and SMDS demographics** Demographic information of program participants.

**"Growing by mentoring: A guide for Data Science for Social Good mentors"** A self-paced learning material for DSSG mentors.

# “Near-peer Mentoring in Data Science: A Plot for Mutual Growth” Supplementary Material

# Data Science for Social Good Program Activities

The Data Science for Social Good program includes several components to support participants' learning and introduce them to the various domain of data science career and research. We describe these activities below and provide an example of weekly calendar of the 2022 program in Table 1. The program was held remotely in 2022, and as a result synchronous meetings are scheduled in the mornings to accommodate different time zones, and participants work asynchronously in the afternoons.

- **DSSG cohort meeting** The entire DSSG cohort meet at the start of each week. Teams present project updates and provide feedback to each other. The head mentor introduces schedules and activities in the coming week. Weekly team meetings facilitate synergy between teams, which is especially important when the program is held remotely, and different teams don't have a structured time to discuss work with each other.
- **Project team meeting** Project team members and technical mentors brainstorm ideas, plan tasks and share updates. The team meeting is typically scheduled at the start of each day to help organizing daily tasks.
- **Meeting with project partners** Project team updates project partner and faculty advisor about their progress once a week. The team works with the project partner to identify directions in the coming week. The project team can schedule additional meetings with the faculty advisor to obtain additional guidance and support.
- **Technical training** Once every two weeks, technical mentors give a tutorial on topics in data science, statistics, or programming. Some examples include "introduction to deep learning", "multiple hypothesis testing", "writing reproducible manuscripts in R", and "how to use high-performance computing". The technical training session provides an opportunity discuss technical topics relevant for all the teams.
- **Data and donuts** Each week, fellows join data science scholars on campus in conversations about data science topics. Each team presents twice during the program on a topic of interest, e.g., "data visualization", "A/B testing", and "fairness in machine learning".
- **Data scientist talks** Mentors invite speakers from academia or industry to discuss their career trajectory and/or research. Previous talks have discussed "Harnessing the power of big team science to discover generalizable insights in the social sciences", "Applying computer vision to remote sensing data for environmental conservation: projects from the AI for Earth initiative", and "Combining mechanistic + ML models in health: When, how, why, and why not?". The talks are in a casual format where students are encouraged to ask questions and interact with the speaker.
- **Visits to external partners** Occasionally, the program organizes visits to industrial partners, where fellows meet data science professionals and learn about their work and career.
- **Social events** Social activities are organized every two weeks where fellows and mentors get to know each other through team building activities, such as improvisation class, virtual painting, and escape rooms.
- **Mentor meeting** Mentors meet with faculty advisors once a week to discuss project progress, activities in the coming weeks, and challenges they have encountered.

**Table 1** An example weekly calendar during the program in 2022. Synchronous portion of the program are scheduled in the morning and participants can work asynchronously in the afternoon.

| Weekday | Time | Activity |
|---|---|---|
| **Monday** | 9-10 AM | DSSG cohort meeting |
| | 10-11 AM | Project team meeting |
| | 12-1 PM | Meeting with project partner |
| | 1-5 PM | Project work |
| **Tuesday** | 9-10 AM | Project team meeting |
| | 12-1 PM | Mentor meeting |
| | 1-5 PM | Project work |
| **Wednesday** | 9-10 AM | Project team meeting |
| | 10-11 AM | Data and Donuts |
| | 1-2 PM | Data scientist talk |
| | 1-5 PM | Project work |
| **Thursday** | 9-10 AM | Project team meeting |
| | 1-5 PM | Project work |
| **Friday** | 9-10 AM | Project team meeting |
| | 11-12 PM | Technical training / Social event |
| | 1-5 PM | Project work |

# Data Science for Social Good

## Project Milestone Template

**Project title:**

**Technical mentor:**

**Faculty mentor:**

**Problem statement:**

**Goal**:

Example:

[1-2 weeks]

1. Organize files
2. Develop a procedure to convert documents into texts.
    1. Convert forms into key-value pairs.
    2. Extract text from narratives

**Data description**: *Please include a high-level description of the available data.*

**Milestones:** *Please fill in the following table. Be as specific as possible.*

| When | What | Details & Comments |
|---|---|---|
| **Week 1** | Example:<br>1. Get to know the project and each other<br>2. Familiarize with project background, understand the big picture<br>3. Brainstorm ideas<br>4. Explore the data<br>5. Prepare for D&D | Example:<br>1. Understand our goal and envision what we want to include in the deliverable<br>2. Study previous data organizing and analysis work<br>3. Data exploration<br>a. Explore the SQL database [SQL tutorial]<br>b. Look at annotated data, summarize your observations |
| | | Each student will pick one of the data and share what they find |
| **Week 2** | | |
| **Week 3** | | |
| **Week 4** | | |
| **Week 5** | | |
| **Week 6** | | **Intermediate presentation to DSSG** |
| **Week 7** | | |
| **Week 8** | | • **Final presentation to DSSG**<br>• **Final code submitted** |

# Example project onboarding material

**Project Background**

- Read about the project background and goals here [link]
- Context
- Previous work

**Statistics**

*Classification methods*

- Assessing classification accuracy, AUC curve
- Fairness in machine learning by Barocas et al. (book chapter [link])

**Matching algorithm**

- Deferred acceptance algorithm: lecture notes [link]
- Related research articles

**Data science in R**

*Getting started*

- Introductory statistics with R by Peter Dalgaar (Chapter [link])
- Introduction to R for python users [link]

*Data visualization*

- Getting started with ggplot2
- Common graphs and how to make them in R

*Working with data*

- Data manipulation
- Joining data tables

**Reproducible data analysis**

- Good enough practices in scientific computing by Wilson et al. [link]

*Github*

- Happy Git and Github for the useR by Bryan et al. [link]
- Tutorials

# Data Science for Social Good Teaching Material Example

**Context**

In the example, participants need to learn to how extract relevant data from a SQL database. To facilitate student learning, the mentor provides tutorials, exercises, and invited a graduate student to give a lecture and answer questions. The participants then apply the newly acquired skill to examine and transform raw data to curate a data set they can work with.

**Tutorial**

Students are given lecture notes and tutorials to use SQL database before an interactive workshop by a graduate student.

1. Lecture notes from one of the classes (we do not share the lecture notes as it is developed by other faculties).
2. A tutorial to access the SQL file used in the project including:
    1. Downloading and installing docker file
    2. Accessing PostgreSQL database
    3. Typing in queries
3. A 1-hour Interactive workshop by a graduate student (we do not share the lecture notes as it is developed by the graduate student)

**Exercise**

After the workshop, students work on a few questions and they input their answers in a shared excel worksheet, letting them see their peer's responses

1. What's your SQL query? (list three)
2. What does your query return?

**Example answers to the exercise**

1. Query: SELECT COUNT(link) from police_records_document This query returns total count of files with links
2. Query: select count(organization_id) from police_records_officer officer where officer.organization_id IN (select organization.id from police_records_organization organization where organization.org= 'A' OR organization.org = 'B' OR organization.org = 'C') This query returns total number of cases in the database from three districts with annotated data

# Stanford Mentoring in Data Science Workshop Topics

1. “Bridges and barriers to belonging” by Marcella Anthony
2. “Diversity and the challenge to academic culture” by Joseph Brown
3. “Navigating mentor-mentee dynamics” by Miranda Stratton
4. “Writing Teaching and Diversity Statements” by Zandra Jordan

# Global Health Activity: R Language

Marissa Reitsma

## Background

Although many people may think of big technology companies when they consider careers in data science, the tools of data science can be applied to just about any field you can imagine. In this example, we'll use the R language to explore global health data on the impacts of different diseases for all countries around the world.

In the United States there is a comprehensive system ("vital registration") that records the cause of death for all individuals. We can use data from the vital registration system to identify the leading causes of death in the US and how these have changed over time as medical advances (prevention, screening, treatment, etc.) have been developed and deployed. For many countries, vital registration systems do not exist, and causes of death are not systematically recorded and compiled. The lack of data presents a major obstacle to improving health. Over the last three decades, increasingly complex statistical models have been developed to estimate causes of death for all countries as part of the Global Burden of Disease Study. These estimates can be used to inform new policies and interventions, resource investments, technology development, and priorities for data collection, among many other applications.

## Exploring an Interactive Data Visualzation

We will first explore the estimates using an online interactive data visualization: http://ihmeuw.org/5f8t. The first view is a square pie-chart, also known as a "treemap." The size of the box corresponds to the percentage of deaths due to a specific cause. Non-communicable diseases are shown in blue. Infectious diseases as well as maternal and neonatal diseases are shown in red. Injuries are shown in green.

1) How have the leading causes of death globally changed between 1990 and 2019?
2) What do you like about the visualization tool? What do you dislike?

## Working with R and Loading Data

Next we will use the R language to explore the tabular data that power the visualization tool. R uses packages to make analyses faster and easier. A collection of packages commonly used for data science in R is Tidyverse . You can install packages in R by naming the package within the install.packages function: `install.packages("tidyverse")`.

You only need to install packages once. After you've installed the package, you can load it into your R session using the library function: `library(tidyverse)`. Loading the tidyverse package loads all of the packages that are part of tidyverse. For now, you'll mainly use dplyr for data manipulation and ggplot2 for making figures.

```r
library(tidyverse)
library(magrittr)
```

Next, load the global health data into your R session. The data is stored in a git repository. Git is a tool for managing code and data, particularly when there are multiple contributors. There is a plenty to learn about git (if you're interested), but for now all you need to do load the data.

```
## Read the data into R using the read_csv function
df <- read_csv("https://raw.githubusercontent.com/conordoherty/BIODS-360-class-materials/master/
               datasets/global%20health/IHME-GBD_2019_DATA-90d5985d-1_clean.csv")
```

Look at the first few rows of data using the head function.

```
head(df)
```

The line reports the dimensions of the data you are viewing: the first six rows of a dataset that has 10 columns. The second line contains the column names. The third line reports the data type of each of the columns: chr stands for character and is a string (ie. text), while dbl stands for double and is a number. You can do math on numbers, but not on strings.

```
sum(df$val) ## This works because `val` is a number
```

```
sum(df$location_name) ## This causes an error because `location_name` is a character
```

You can change the number of rows you're looking at by passing an additional argument to the head function. You can learn more about what arguments exist for a function by using ? before a function in R (or trusty Google!): `?head`

```
head(df, n = 10)
```

## Analyzing Data

There are endless questions to answer with these data. There are also many ways to get to one answer using various functions in R. We can learn some of the functions by answering different questions in our dataset.

### Subset and Group

We'll learn the subset and group functions by answering the following questions:

1.) How many deaths occurred globally in 1990 versus 2019?

2.) How many deaths occurred in the United States in 2019?

```
## How many deaths occurred globally in 1990 versus 2019, by sex and for both sexes combined?
df %>%
  group_by(year, sex_name) %>% ## Summarize separately by year and sex
  summarise(total = sum(val)) ## Report the sum of the column "val" (which is the mean estimate)
```

```
## How many deaths occurred in the United States in 2019?
df %>%
  ## Select only rows where location_name == "United States of America"
  subset(location_name == "United States of America" & year==2019 & sex_name == "Both") %>%
  summarise(total = sum(val)) ## Report the sum of the column "val" (which is the mean estimate)
```

**Sort**

We'll learn the arrange function, which sorts rows by values in a specified column, by answering the following question:

1.) What were the leading causes of death in the United States in 2019?

```
## What were the leading causes of death in the United States in 2019?
df %>%
  ## Select only rows where location_name == "United States of America"
  subset(location_name == "United States of America" & year==2019 & sex_name == "Both") %>%
  arrange(desc(val)) %>% ## Report the sum of the column "val" (which is the mean estimate)
  ## Compute a new column that reports thousands of deaths, rounded to two decimal places
  mutate(deaths_thousands = round(val/1000, 2)) %>%
  select(-c("val", "upper", "lower", "metric_name", "age_name")) %>% ## Drop unnecessary columns
  head(n=20) ## Show the first 20 sorted rows
```

## Visualizing Data

Visuals are powerful tools to convey information to a variety of audiences. The `ggplot2` package in R provides a flexible approach to generating beautiful figures that are built with a consistent set of inputs: data, a coordinate system, and "geoms." Geoms are visual representations of the data, such as points, lines, and bars.

We'll learn the basics of ggplot by answering the following question:

1.) How have the top 10 causes of death in the United States changed between 1990 and 2019?

First prepare the data that will be used in the figure:

```
## Using the skills we learned in the previous section,
make a new object called "plot data" that contains the summarized data we want to plot
plot_data <- df %>%
  ## Select only rows where location_name == "United States of America"
  subset(location_name == "United States of America"  & sex_name == "Both") %>%
  select(location_name, cause_name, year, val) %>%
  spread(year, val) %>%
  arrange(desc(`2019`)) %>% ## Report the sum of the column "val" (which is the mean estimate)
  ## Compute a new column that reports thousands of deaths, rounded to two decimal places
  mutate(`1990` = round(`1990`/1000, 2)) %>%
  ## Compute a new column that reports thousands of deaths, rounded to two decimal places
  mutate(`2019` = round(`2019`/1000, 2)) %>%
  head(n = 10) %>%
  gather(year, deaths_thousands, 3:4)
```

Then make a basic figure:

```
## Make a bar chart using ggplot
ggplot(data = plot_data, aes(x = cause_name, y = deaths_thousands, fill = year)) +
  geom_bar(stat = "identity", position = "dodge")
```

That looks pretty bad, but all the pieces are there. We'll want to improve the readability of the labels and also rotate the x-axis labels to avoid the text overlap:

```r
## Make a bar chart using ggplot, add proper labels
ggplot(data = plot_data, aes(x = cause_name, y = deaths_thousands, fill = year)) +
  geom_bar(stat = "identity", position = "dodge") +
  labs(x = "Cause of Death", y = "Deaths (Thousands)",
       title = "Change in Top 10 Leading Causes of Death in the US, 1990-2019", fill = "Year") +
  theme_bw() +
  theme(axis.text.x = element_text(angle = 70, hjust = 1))
```

Last tweak is to order the bars by number of deaths:

```r
## Make a bar chart using ggplot, add proper labels, order by number of deaths
ggplot(data = plot_data, aes(x = reorder(cause_name, -deaths_thousands),
                             y = deaths_thousands, fill = year)) +
  geom_bar(stat = "identity", position = "dodge") +
  labs(x = "Cause of Death", y = "Deaths (Thousands)",
       title = "Change in Top 10 Leading Causes of Death in the US, 1990-2019", fill = "Year") +
  theme_bw() +
  theme(axis.text.x = element_text(angle = 70, hjust = 1)) +
  theme(legend.position = "bottom")
```

Try using the skills that you've learned to answer your own questions! If you have questions about the data, feel free to contact Marissa Reitsma mreitsma@stanford.edu.

# DSSG and SMDS Program Demographics

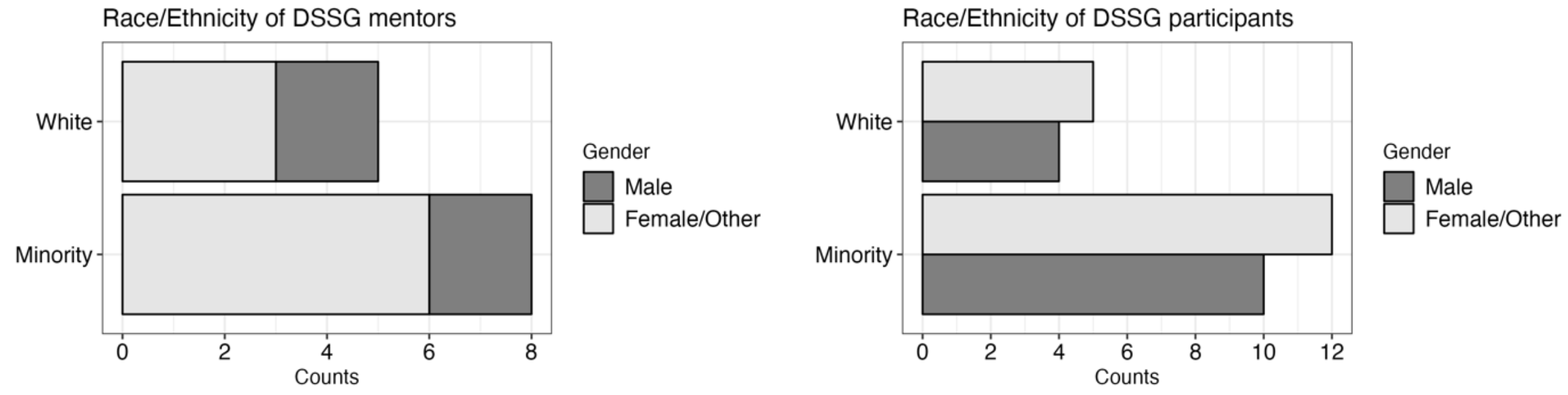

**Figure 1** Data Science for Social Good participant and mentor demographics from 2019 -- 2022. The racial minority groups include individuals who identify themselves as American Indian and Alaska Native, Asian, Black, Hispanic, and Native Hawaiian/Pacific Islander.

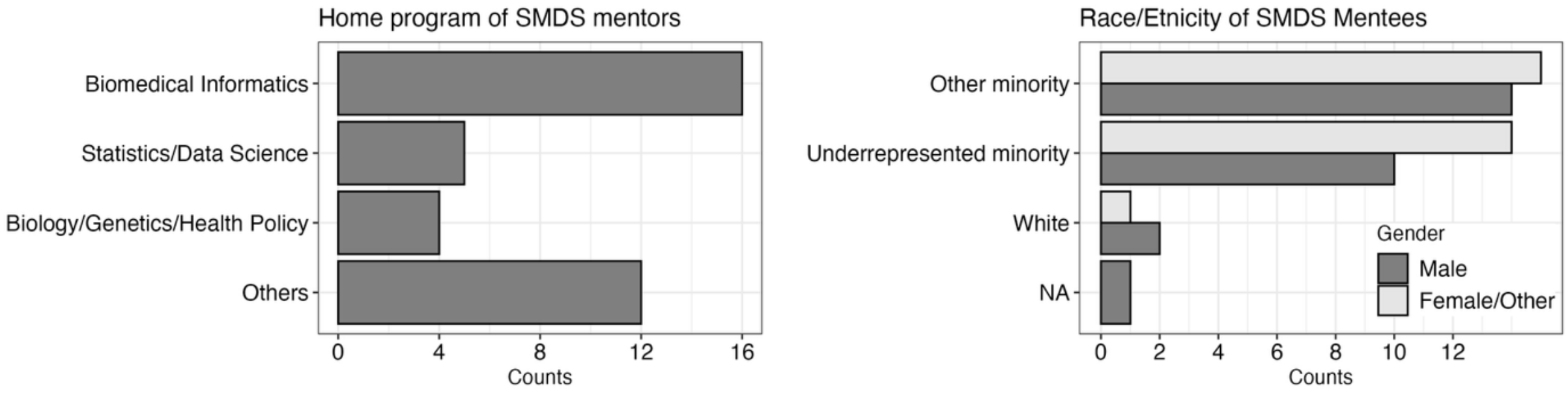

**Figure 2** Demographic information of Stanford Mentoring in Data Science participants and degree program of mentors. We use the definition of underrepresented minority groups in science and engineering professions (blacks or African Americans, Hispanics or Latinos, and American Indians or Alaska Natives) by National Center for Science and Engineering Statistics (NCSES). Other minority groups are defined as minority groups defined by NCSES that are not one of the underrepresented minority groups (Native Hawaiians or Other Pacific Islanders, Asians, and individuals reporting more than one race). Mentors come from a wide range of academic programs, including (but not limited to) civil and environmental engineering, economics, computer science, management science and engineering, and earth system science. A mentor is counted multiple times if they have participated several times in the program.

Qian Zhao

*Growing by Mentoring:*
*A Guide for Data Science for Social Good Mentors*

Stanford

# *Contents*

# *Introduction*

Welcome, Data Science for Social Good mentors!

We are excited that you are committed to spending the next few months mentoring students on a data science project with social impact. Your guidance are crucial to the success of the project and your students' growth as a data scientist. Besides being a learning experience for the fellows and enhancing data analysis capacity for the project partner, Data Science for Social Good program (DSSG) also provides an opportunity for you to grow as a mentor, teacher and project manager.

If this is your first DSSG experience, you can begin by going through one the following projects to get an idea of what a typical project looks like. Throughout the guide you will encounter many "activities", where we invite you to reflect on your own experiences or design materials you can use during DSSG. By the end of this guide, you will have a concrete list of materials, e.g., a learning activity for a tutorial, a project milestone, and learning goals for fellows. These activities are self-paced, and you can complete these activities on your own or with others.

**Activity** Go through the following DSSG projects, what are the goals of each project? What approaches did the team take? How did they present their results?

1. Identifying Factors Driving School Dropout and Improving the Impact of Social Programs in El Salvador [1]

2. Predictive Enforcement of Pollution and Hazardous Waste Violations in New York State[2]

3. Developing NLP Tools for Sharing of Indigenous and Community Knowledge.[3]

[0] Tjiptarto, D., Valdivia, A., Baltasar del Valle-Inclan Redondo. (2018). *Identifying Factors Driving School Dropout and Improving the Impact of Social Programs in El Salvador* Data Science for Social Good, Carnegie Mellon Unviersity
[0] Jin, J., Kamenetsky, M., Magee, D. (2016). *Predictive Enforcement of Pollution and Hazardous Waste Violations in New York State* Data Science for Social Good, Carnegie Mellon Unviersity.
[3] Chow, A., Dong, H.W., & Kuang, Y. Q. (2019). *Developing Natural Language Processing-Tools for Sharing of Indigenous andCommunity Knowledge* Data Science for Social Good, University of British Columbia

You might notice that these projects span many application domains, e.g., education equity, environment protection, and energy industry. Their goals range from exploratory data analysis to decision making. The DSSG teams use a variety of tools, such as data visualization and predictive modeling. Nonetheless, the projects share several common features. For instance, they have clear problem statements and impacts; they involve steps in a typical data science project cycle, such as exploratory analysis and feature engineering; the deliverables often include a public presentation to showcase the project, a written report to document the analysis, and an organized Github repository to reproduce the results.

I prepared this guide as a starting point for you to design your mentoring experience. Because this guide is based on reflections of my experience as a fellow and technical mentor at Stanford DSSG summer program, I wrote this with the Stanford program as the point of reference. At the same time, I include general topics and examples from several DSSG programs, so that I hope you find the guide helpful in your program as well.

The grounding point of this guide is the *growth mindset*, the belief that (1) one's intelligence and ability develops through continuous effort, (2) challenges present opportunities for learning, and (3) we can be inspired by others' success[4]. With this in mind, this guide is composed of three sections corresponding to different areas of growth:

[4] Carol S. Dweck. *Mindset: The New Psychology of Success.* Random House, 2006

- *Your growth as a mentor* I invite you to reflect on what good men-

tors *are* and what they *do*. This section will guide you to set goals for your experience, and provide strategies of inclusive mentoring.

- *Development of your project* This section focuses on how to manage a complex project and how to lead a team of fellows. I provide strategies to collaborate with a community partner and to facilitate team work.

- *Fellow's growth as data scientists* In this section, I invite you to think about what learning means. I will guide you to come up with concrete steps to help fellows improve their data science skills, to reflect on their own learning and work, and to help each other grow as data scientists.

In each section, I will explain relevant concepts, discuss strategies, and provide resources where you can learn more about each topic. The materials are designed such that you can read the sections in orders you prefer.

Before we start, I invite you to familiarize with your program structure.

> **Activity** Read your program schedule, what are the components of the program? Who will participate? What are your roles?

For example, Stanford DSSG program includes the following components (see Figure 1 for an example of weekly activities):

- *Weekly group meeting* Fellows present project updates and discuss activities in the past and coming week.

- *Daily team meeting* Project team members and technical mentor(s) brainstorm ideas, plan tasks and share updates with each other.

- *Data and donuts* Fellows join Stanford data science scholars in conversation about data science topics, e.g., data visualization, A/B testing and fairness in machine learning.

- *Technical training* Technical mentors give a tutorial on data science topics, e.g., the basics of deep learning, multiple hypothesis testing, and writing reproducible manuscripts in R.

- *Data scientist talks* Speakers from academia or industry discuss their career trajectory and research topic. Topics include "Harnessing the power of big team science to discover generalizable insights in the social sciences", "Applying computer vision to remote sensing data for environmental conservation: projects from the AI for Earth initiative", and "Combining mechanistic + ML models in health: When, how, why, and why not?".

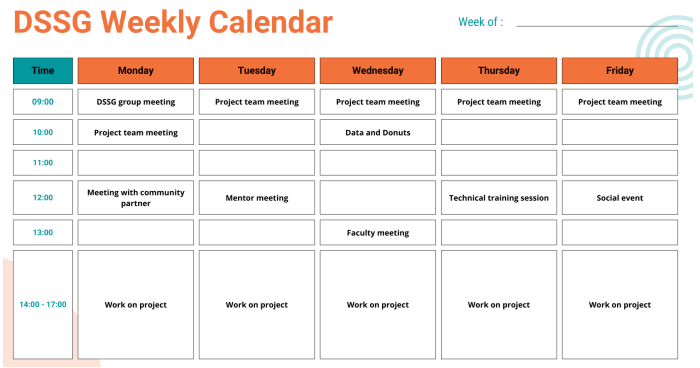
DSSG Weekly Calendar

Week of :

| Time | Monday | Tuesday | Wednesday | Thursday | Friday |
|---|---|---|---|---|---|
| 09:00 | DSSG group meeting | Project team meeting | Project team meeting | Project team meeting | Project team meeting |
| 10:00 | Project team meeting | | Data and Donuts | | |
| 11:00 | | | | | |
| 12:00 | Meeting with community partner | Mentor meeting | | Technical training session | Social event |
| 13:00 | | | Faculty meeting | | |
| 14:00 - 17:00 | Work on project | Work on project | Work on project | Work on project | Work on project |

Figure 1: Example of weekly activities at Stanford DSSG program.

- *Visits to external partners* Fellows and mentors visit industrial partners, meeting data science professionals and learning about their work.
- *Social events* Fellows and mentors get to know each other through social activities, such as improvisation class, virtual painting, and escape room.
- *Mentor meeting* Mentors meet with organizers and faculty advisors to discuss project progress, make plans for the coming weeks, and address issues.

These events create opportunities for fellows to learn from each other, to broaden fellows' perspective about various data science applications, and encourage different perspectives on a project. Fellows are typically upper undergraduate or master's students, who have studied fundamentals in computing, statistics and software, but may not have worked on a real world problem. The project team is mentored by technical mentors and a faculty advisor, and the project is usually a collaboration with a partnering organization. The program is organized graduate student organizers and faculty advisors.

# 1
# *Your growth as a mentor*

As you will be interacting with fellows regularly for a few months, DSSG provides an excellent opportunity for you to experience mentoring, whether you have served as a mentor before or not. In this section, you will reflect on qualities and behaviors of good mentors. You will set your mentoring goals as well as set goals with DSSG fellows. I will discuss two specific topics in mentoring: belonging and how to create an inclusive learning environment. Since I will ask you to set goals many times throughout the guide, I will first discuss why we set goals and what are good goals.

## *1.1 Why do we set goals and what are good goals?*

I typically enjoy a trip more if I have planned a few stops beforehand. Even if I do not end up at the final planned destination or take occasional detours, having a few stops in mind allows me to plan my route and prepare for the trip. This example speaks to one benefit of goal setting: it provides direction and orients us towards relevant activities. Like Seneca said, "When a man does not know what harbor he is making for, no wind is the right wind." In contrast, once goals are clear, one can develop strategies and evaluate their progress.

You might still wonder, are goals equally effective? What are good goals versus bad goals? Here, I refer you to two frameworks that categorizes good goals. The first one is conceived by Latham and Locke. In this framework, a good goal should be both *specific* and *challenging*, because a goal that is both clear and stimulating tends to motivate people to exert more effort and to persist[1]. At the same time, the goal should be *manageable*, i.e., it should be perceived as attainable, and necessary resources and support should be available. One should also be *committed* to the goal and routinely analyze their progress or receive *feedback* from others, since they can adjust your both objective and approach accordingly and thus improve their performance.

[1] Gary P. Latham Edwin A. Locke. *A Theory of Goal Setting & Task Performance*. 1990

Another goal setting framework was developed by Doran from the perspective of management. A good goal in this framework should be *Specific, Measurable, Achievable, Realistic* and *Time-oriented*[2] (S.M.A.R.T. in short). When you write down a goal, you can ask a few questions: is this goal relevant for the long-term objective? Is there a time-frame for when I should plan to achieve the goal? What skill do I need and what resources are available? When should I re-evaluate my goal based on my progress? Though you should not modify your goals frequently, you can re-evaluate them, and the purpose of setting goals is provide a signpost with which you can measure how far you have travelled.

[2]

## 1.2 *Setting your mentoring goals*

**Activity** Think about a wonderful mentor you have had in the past: what are some of their qualities? What did they do that were most helpful in your (1) scientific and (2) personal development? [1]

[1] These questions are inspired from questions in the paper *The Qualities and Impacts of a Great Mentor—and How to Improve your own Mentoring* by Grogan,P., and Eviner, V., & Hobbie, S. (2013) in *The Bulletin of the Ecological Society of America*.

Perhaps your mentor was warm, supportive and positive; perhaps they suggested ideas when you were stuck on your research; perhaps they were your academic role model. All of these are important qualities. Let's say you encounter a roadblock in your project, by being positive your mentor shows that they accept you as a young scholar, and that they are confident in your ability to address the challenge. By being supportive, they encourage you to become an independent researcher and to push the boundary of knowledge in your domain. By offering guidance, they help you move forward in your work. These qualities help you develop personally and professionally, both of which are important aspects in mentoring.

### 1.2.1 *What is mentoring?*

While we may consider our peers who demonstrate academic excellence or family member who give us suggestions as our mentor, in this guide I focus on mentoring in a *formal* and *academic* setting. A mentoring relationship is formal when a mentee is assigned to the mentor, and the mentoring relationship has certain structure, length, and guideline[3]. I also focus on mentoring in the academic setting when mentors offer knowledge, advice, challenge, counsel and support in the protégé's pursuit of being a full member of a profession[4].

I adopt the definition of mentoring in Packard (2016)[5]:

[3] Lillian T. Eby, Jean E. Rhodes, and Tammy D. Allen. *Definition and Evolution of Mentoring*, chapter 2, pages 7–20. John Wiley & Sons, Ltd, 2007

[4] W. Brad Johnson. The intentional mentor: Strategies and guidelines for the practice of mentoring. *Prof Psychol Res Pr*, 33:88–96, 2002

[5] Becky Wai-Ling Packard. *Successful STEM Mentoring Initiatives for Underrepresented Students : A Research-Based Guide for Faculty and Administrators*. Stylus Publishing, 1st edition, 2016. ISBN 9781620362952

> Mentoring (is) a developmental experience or a type of support intended to advance students toward an important goal.

This definition highlights three aspects of mentoring:

- Mentoring is *student-oriented*, and the objective is to guide mentee towards accomplishing a goal. Mentoring starts with understanding needs and aspirations of the mentee. In the next section, you will see a few questions to help you understand fellows' academic journey, goals, and interests.
- While the "developmental experience" refers primarily to the growth of the mentee, mentors tend to also benefit from the mentoring relationship. As Galbraith pointed out, "mentoring a powerful and passionate interaction, whereby the mentor and protégé experience personal, professional, and intellectual growth and development"'[6].
- Mentoring is a reciprocal relationship that evolves through time[7]. Thus, characteristics of a nurturing relationship (e.g., open, respectful, and authentic) make for more effective mentoring.

While each individual has their unique mentoring style, mentees typically appreciate certain attributes and conducts of of their mentors. The article *Nature's guide for mentors*[8] quotes the description of several scientists' of their mentors. The following are a few excerpts from the article:

- *Active listening and questioning* "There is always another question to ask. The questions seem innocuous but nothing is as it seems to be; there are more insights to be gained by probing away. M also never imposes her will, but she persistently keeps the questions flowing to help the answer come along."
- *Balancing direction and self-direction* "The major aspects of practice and personality are her ability to listen patiently, even when she knows better, and to point the mentored person to a more complete understanding of the issues implicit in a particular problem. This she does with deceptively simple questions that frequently do not elicit an immediate response, but ultimately allow a more rational interpretation of all the facts."
- *Skill development* "M has focused on equipping people with the skills to be fully functioning members of the scientific community, able to prepare grant applications, review manuscripts, speak at conferences and engage with scientific administrators in a constructive manner. Such a holistic approach to running a scientific group will ultimately bring enormous benefit to the group's

[6] Michael W. Galbraith. Celebrating mentoring. *Adult Learning*, 14(1):2–3, 2003

[7] Joyce Fletcher and Belle Ragins. *Stone Center Relational Cultural Theory: A Window on Relational Mentoring*, chapter 15, pages 373–399. Los Angeles: Sage Publications, 2007

[8] Adrian Lee, Carina Dennis, and Philip Campbell. Nature's guide for mentors. *Nature*, 447(7146):791–797, 2007

alumni, giving them all the skills necessary to carve out their own niches in the academic world."

**Activity** Read the article *Nature's guide for mentors*. Which description do you find particularly inspiring? How does the mentor's behavior assist the personal or professional development of the mentee? Then, Thank about one or two things that you can do as a mentor. You can use Table 1 in the paper as a guide.

Now that you have seen what good mentors *do*, I retrace to the beginning of this section where you reflected on the qualities of a wonderful mentor you have had. What do you want to *be* to your mentees? What qualities do you want to demonstrate?

To help you formulate an answer, I introduce the *tiangular model of mentor competence* from the book *On being a mentor*[9]. This framework considers three aspects of mentor competency: virtues, abilities and competency. *Nature's guide to mentors* provides many examples of competencies, and you can find more examples in Chapter 5 of the book, and now let us focus more on virtues and abilities.

[9] W. Brad Johnson. *On being a mentor: A guide for higher education faculty.* Lawrence Erlbaum Associates Publishers., 2007

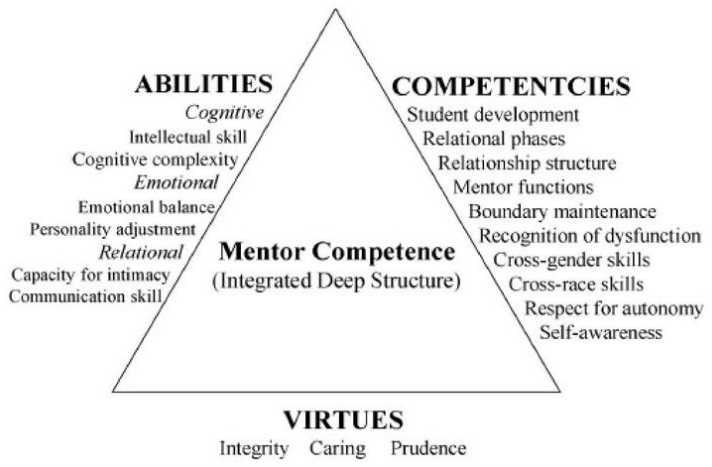

I highlight two mentoring abilities from the triangular framework: emotional ability and relational ability. Emotional ability speaks to the capability to understand and manage one's emotion in life and relationships. One example of emotional ability is empathy, which is the ability and willingness to listen, understand, and reflect on others' needs and concerns. Another example is being friendly and collegial, showing genuine respect and acceptance of someone's individuality, skill and experience. An example of relational ability is congruence, where a mentor accurately and authentically communicate their experience with the mentee. Relational abilities are important because by exhibiting excellent interpersonal skills, mentors help their mentees develop their interpersonal skills as well.

We emphasize that these positive attributes can be cultivated. For instance, you may not think of yourself as a particularly warm person, and yet you can still make fellows feel welcome through your actions. To conclude this section, I invite you to reflect on how you can apply some of these concepts in your mentoring.

**Activity** What are your strengths and weaknesses as a mentor? Which aspects do you want to grow in? What goals do you want to set for yourself? What steps do you plan to take to make progress in these directions? (e.g. I can compliment students when they ask good questions, make good comments or have good ideas. For example, I can tell them "That's a good idea! I did not think of it!")

## 1.3 Getting to know DSSG fellows

Getting to know your students not only help you connect with them, but also it allows you to tailor your mentoring based on their needs. The practice of determining the mentoring outcomes before starting a program and then plan the course of action accordingly is called *intentional mentoring*[10]. As you can imagine, we can help students progress towards their goals more effectively by being intentional.

[10] Becky Wai-Ling Packard. *Successful STEM Mentoring Initiatives for Underrepresented Students : A Research-Based Guide for Faculty and Administrators*. Stylus Publishing, 1st edition, 2016. ISBN 9781620362952

### 1.3.1 Skills fellows can learn through DSSG

DSSG fellows are typically come from various personal and academic backgrounds. They also tend to have different motivations and goals. For instance, some of them might want to enrich their professional portfolio, others may want to be exposed to academic research; some may be interested in educational equity while others may be curious about epidemiology. Nevertheless, they all want to become more proficient data scientists, and the multifaceted activities during DSSG provide ample opportunities for fellows to grow.

The following a list of data science skills fellows can enhance through DSSG. You can use this as a reference when you think about directions your mentees might want to pursue. The list, based on *The data for good growth map*[11] and *Curriculum guidelines for undergraduate programs in statistical science*[12], is not meant to be comprehensive, and different skills are grouped to ease your reading.

[11] Dharma Dailey, Sarah Stone, Anissa Tanweer, and the Data for Good Organizer Network. The data for good growth map: Decision points for designing a university-based data for good program

[12] American Statistical Association Undergraduate Guidelines Workshop. Curriculum guidelines for undergraduate programs in statistical science, 11

- Contextual knowledge
  - Understanding background and purpose of a project, e.g., by reading literature and communicating with the project partner.
  - Learning domain knowledge for a specific subject and engaging with domain experts.
  - Creating a deliverable and presentation tailored to the project partner's needs.
- Collaboration skills
  - Working with team members.
  - Providing feedback to each other.
- Technical research skills
  - Managing a data set, e.g., identifying missing data, data transformation and encoding.
  - Visualizing data, e.g., emphasizing relations between variables of interest, leveraging graphical elements to facilitate comparison between different pieces of data.

- Modeling, including model selection, evaluation and communication.
  - Conducting reproducible research.
  - Using collaboration tools, e.g., Git.
  - Becoming more fluent with statistical software, e.g., python or R.
  - Designing and implementing simulations or optimizations.
  - Developing mathematical understanding of a statistical model.

- Becoming a self-directed learner

  - Planning, carrying out and evaluating a data analysis approach.
  - Identifying knowledge gap and resources.

- Career development

  - Familiarizing with data science project cycle.
  - Managing a project.
  - Introducing themselves to community partners.
  - Communicating their research with other fellows, faculties and public.
  - Networking with data science professionals.

### 1.3.2 *Setting goals with fellows*

Now that you have seen what skills fellows can sharpen, I invite you to set personal development goal with fellows. You can use the worksheet in Table 1.1. These questions are designed to help you understand their academic journey and their professional interest. Goal setting is an *iterative* process, and I encourage you to meet with fellows regularly, and modify your answer throughout the program.

Table 1.1: Fellows goal-setting worksheet

1. What has been your academic journey?
2. What are some interesting projects you have worked on or courses you have taken?
3. What's your academic goal in the next year (what are your long term goals)?
4. How can DSSG help you achieve your goals?
5. What technical skills do you want to improve?
6. Other questions about the fellow's interest or value, for example, are they interested in educational equity, environmental justice etc.?

**Activity**

1. Meet with your mentees (you can use Table 1.1 as a reference); take notes from the meeting.

2. After the meeting, think about your *big dream* for your mentees: What do you want to them to achieve after the DSSG program and one year after the program? With this vision in mind, write down one or two things you or the fellow can do to accomplish this. For example, if a fellow wants to improve their presentation skills, then you can spend some time to practice presentation with them. You can also encourage them to take the lead on presenting at one of the group meetings. If a fellow is interested in developing their professional portfolio, then you can connect them with data scientists, invite speakers to talk about their career trajectory, or provide resources to help them navigate data science job market.

## 1.4 *Fostering a sense of Belonging*

In 1967, Jocelyn Bell Burnell noticed a strange pulsing signal from the radio telescope. The signal did not look like that from a quaser, a type of active galactic nucleus with a massive blackhole in the center surrounded by an accretion disk, and neither was it the result of interference. “Immediately, I knew it was something else”. She took note of when the time the signal is expected to occur, and switched the machine to high speed setting to catch its detail. It was a string of pulses, one and a third seconds apart from each other. Later, she observed a second and a third of this signal, confirming her idea that this belongs to a new object. She had discovered *pulsar*, the pulsating radio star, which is the remnants of big star that spins and emits radio signals. The discovery would later earn her advisor Antony Hewish the Nobel prize even though she was the one who discovered and investigated the anomalous signal.The Nobel prize in physics in 1974 was awared to Antony Hewish, for his decisive role in the discovery of pulsars, and Martin Ryle, for his observations and inventions, in particular of the aperture synthesis technique.

In several interviews, Jocelyn Bell Brunell said she has been motivated by her feeling as outsider and her imposter syndrome.“They are going to discover their mistake (of admitting me) and throw me out. I got around it by deciding I would work my very hardest, so that when they threw me out I wounldn’t have a guilty conscience”. She also described her experience in her undergraduate studies as the only girl in a class of fifty. “It was the tradition, when a women entered the lecture, all the male students stamped and whistled and called and banged the desk. I had to face this on my own. It's nasty, yes. If I hadn't been clear what I wanted to do, I would have gone some other way."[13]

While Jocelyn Bell Burnell persisted through the scoffing, the feeling of not belong is one of the key factors that turn underrepresented students away from the academia. As members of the data science community, we all have parts to play to ensure everyone feel that they belong. In this section, I will discuss some challenges for minority students in the academic setting and what are some strategies you can take to create an inclusive learning environment where everyone can thrive.

[13] Ben Proudfoot. She changed astronomy forever. he won the nobel prize for it. https://www.nytimes.com/2021/07/27/opinion/pulsars-jocelyn-bell-burnell-astronomy.html, year = 2021. New York Times, Op-Docs

### 1.4.1 *What is belonging?*

Belonging is the perception of "being connected to others: accepted and included, to be valued members of social groups, and to contribute positively to the lives of others"[14]. Belonging should not require one to make efforts to "fit in", hiding parts of their identity. Rather, belonging means that every aspect of one's identity is respected and valued. In the academic setting, belonging takes additional significance: it means that one feels that they are a member of a profession or a community of practice (e.g.,"I am a data scientist"),[15] that their contributions are valued, and they are respected personally and recognized professionally[16].

[14] Gregory M. Walton and Timothy D. Wilson. Wise interventions: Psychological remedies for social and personal problems. *Psychol. Rev.*, 125:617, 10 2018b

[15] A community of practice is defined by Wenger as "groups of people who share a concern or a passion for something they do and learn how to do it better as they interact regularly" .

Etienne Wenger. Communities of practice: A brief introduction. https://www.nytimes.com/2021/07/27/opinion/pulsars-jocelyn-bell-burnell-astronomy.html, year = 2011

[16] Communications Team Royal Society of Chemistry. A sense of belonging in the chemical sciences

| |
|---|
| **Activity** Think about a situation when you feel that you did not belong in a scholarly environment, what factors contributed to your feeling? What support did you receive or wish you had received? |

You might notice that when you did not feel you belong, you were anxious, more stressed and deterred from pursuing opportunities. In fact, Fisher et al. (2019) showed that while women and African American graduate students tend to publish less than their majority peers, they are more likely to publish at a comparable rate when they feel they are well-prepared for the program, accepted by their colleagues, and experience well-structured programs[17]. You might also notice that the sense of belonging can be affected by small gestures. For example, I felt out of place when one student refused to sit next to me during lunch. In another occasion, I felt that I belong when a professor kindly asked my opinion during a meeting and told me that they were interested to hear my thoughts. Indeed, small gestures make a big impact!

[17] Aaron J Fisher, Rodolfo Mendoza-Denton, Colette Patt, Ira Young, Andrew Eppig, Robin L Garrell, Douglas C Rees, Tenea W Nelson, and Mark A Richards. Structure and belonging: Pathways to success for underrepresented minority and women phd students in stem fields. *PloS one*, 2019

### 1.4.2 *Why is belonging important?*

Many psychologists and sociologists have recognized belonging as one of our basic social needs. In Maslow's hierarchy of human needs, "belonging and love" is considered a *deficiency need* that should to be met before one can fulfull one's *growth needs*, such as the need to know, explore and achieve, and the need to fulfill one's potential.[18] Through the lens of sociobiological theory, Nohria, Lawrence, and Wilson (2001) also recognize one's need to bond with others in long-term relationships of mutual care and commitment [19].

[18] Maslow first formulated his hierarchy of needs on 1954, and later in 1971 he refined the categories in self-actualization needs.

[19] William G. Huitt. Maslow's hierarchy of needs. educational psychology interactive, 2007. Educational Psychology Interactive

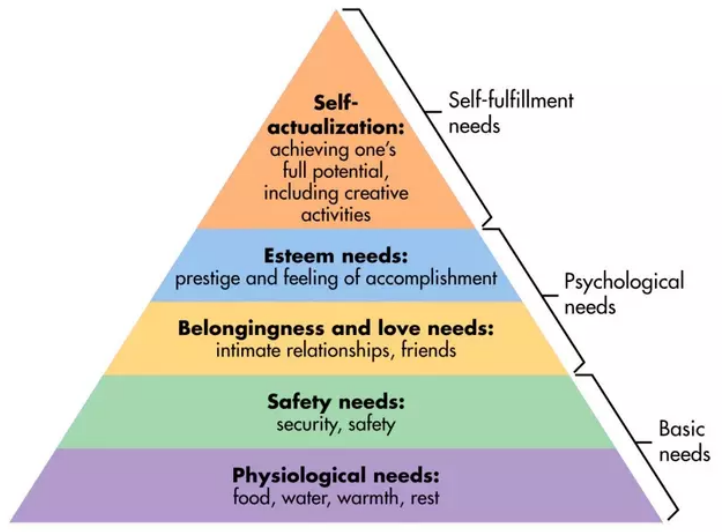

Figure 1.1: Maslow's hierarchy of needs.

In a learning environment, a higher sense of belonging enhances motivation, engagement and persistence. On the other hand, the feeling that one does not belong increases anxiety and depresses learning and performance[20] [21]. In the long term, belonging affects a student's decision to persist in an academic field. One research finds that among college students who leave STEM majors, 70% feel that they did not belong, whereas only 10% feel that they do not belong among students who stayed in STEM majors[22]. The same study also finds that female and minority students are more likely to feel that they do not belong. Nilanjana Dasgupta summarized succinctly in her interview with NSF: "Usually, people walk through the door if they have some degree of ability, interest, and curiosity about a subject. What makes them stay is belonging."[23]

The research suggests that students need to feel they belong regardless of where they are in their academic journey. Moreover, we should create a welcoming environment not only in the formal setting like a classroom, but also in day-to-day interactions. Next, I will describe a few strategies you can use to foster a sense of belonging.

[20] D. L. Schwartz, J. M. Tsang, and K. P. Blair. *The ABCs of How We Learn: 26 Scientifically Proven Approaches, How They Work, and When to Use Them.* W W Norton & Co., 2016

[21] Vincent Tinto. Classrooms as communities: Exploring the educational character of student persistence. *J Higher Educ*, 68(6):599–623, 1997

[22] Katherine Rainey, Melissa Dancy, Roslyn Mickelson, Elizabeth Stearns, and Stephanie Moller. Race and gender differences in how sense of belonging influences decisions to major in stem. *Int. J. STEM Educ.*, 5(1):10, 2018

[23] NSF Research News. 'Belonging' can help keep talented female students in STEM classes, 2016. URL `https://beta.nsf.gov/news/belonging-can-help-keep-talented-female-students-s`

### 1.4.3 *Mentoring strategies to enhance belonging*

**Activity** Reflect on a time when someone made you feel that you belong in a scholarly setting, what did they do?

Perhaps they told you failure is part of the learning process and to cheered you on when you were struggling, perhaps they encouraged you to apply to a professional conference, or perhaps they took the time to greet you when you meet in the hallway. These small actions show that they value your growth, respect acknowledge your scholarly contribution, and respect you as an individual. Below are a few strategies that I found particularly useful, and I suggest you read the article *Ten simple rules for supporting historically underrepresented students in science*[24] if you would like to learn more practical tips.

[24] Suchinta Arif, Melanie Duc Bo Massey, Natalie Klinard, Julie Charbonneau, Loay Jabre, Ana Barbosa Martins, Danielle Gaitor, Rhiannon Kirton, Catalina Albury, and Karma Nanglu. Ten simple rules for supporting historically underrepresented students in science. *PLoS Comput. Biol.*, 17(9):1–16, 09 2021

*Getting to know each fellow* One needs to be seen, respected and accepted to feel that they belong. This means that you should connect with fellows individually, understanding their unique circumstances, academic journey and aspirations. You should take their stories seriously, listen without judgement, and treat them with respect[25]. During your conversations, you should also try to be aware of your own bias or prejudice, judgemental attitudes or behavior, and avoid remarks that convey stereotypes.

[25] Gail Horowitz. *Teaching STEM to first generation college students : a guidebook for faculty & future faculty*. Charlotte, NC : Information Age Publishing, Inc., 2019

**Examples**

- Suppose you notice someone repeatedly using headphones during class in spite of your explicit instruction. Instead of telling them to to remove their headphones, you can try to understand *why* they need to use them. They may have attention deficit hyperactivity disorder (ADHD), in which case background white noise is beneficial for them to listen and read[26]. In the setting of DSSG program, suppose that you find a fellow's behavior frustrating or disruptive, rather than accusing or insisting them to change their behavior, you can discuss your point of view while being understanding and keeping an open mind.

[26] Suchinta Arif, Melanie Duc Bo Massey, Natalie Klinard, Julie Charbonneau, Loay Jabre, Ana Barbosa Martins, Danielle Gaitor, Rhiannon Kirton, Catalina Albury, and Karma Nanglu. Ten simple rules for supporting historically underrepresented students in science. *PLoS Comput. Biol.*, 17(9):1–16, 09 2021

- Take the time and effort to connect with your mentees, e.g., you can arrive a few minutes early before each session to talk to available fellows, or arrange to meet with each of them. This demonstrates that you care about their growth and motivates them to learn because it shows that their learning matters[27].

[27] Lang James M. *Small Teaching : Everyday Lessons From the Science of Learning.*, volume 2nd. Jossey-Bass, 2021

*Reframing challenges* You are probability familiar with failures and consider it part of the scientific process. However, fellows who do not have previous research experience might interpret struggling in their project as a sign that they are not good at data science, and this may trigger stress and anxiety. When you notice that a fellow is frustrated, you can help them *re-frame* the challenge as a valuable learning experience, and convey your confidence in their ability to overcome it.

The act of re-framing is important because while our interpretation of our social environment is relatively fixed, it is at the same time amenable. Like a working hypothesis, we continuously modify our perception a situation. As our interpretation of a situation changes, we also change our behavior and thus has the potential to change the outcome[28]. In an academic study, researchers divided college freshmen into two groups. The students in the treatment group commits to being exposed to the possibility of increased belonging. When college starts, they were asked to read counterfeit survey results from

[28] Gregory Walton and Timothy Wilson. Wise interventions: psychological remedies for social and personal problems. *Psychological Review*, 125:617–655, 2018a

senior students, which indicated that most senior students were worried about if they belonged at the start of college, but their sense of belonging gradually increased as their studies went on. Freshmen were also asked to reflect on how their own experiences echoed the survey and recorded their reflections in a video. As for students in the control group, the survey and essay topics are unrelated to belonging. The researchers collected student GPA at the start and end of college, and they found that while the intervention does not affect the GPA of European Americans, African American students in the treatment group sees a much higher improvement in GPA compared to students in the control group, decreasing the achievement gap compared to European Americans by 79%![29]

In this study, the counterfeit survey suggests that setbacks, rather than indicating that a student does not belong, are common and can be overcome by persistence. As a consequence, students were able to perceive challenges in a neutral light, and thus they feel less stress and perform better.

[29] Gregory M. Walton and Geoffrey L. Cohen. A brief social-belonging intervention improves academic and health outcomes of minority students. *Science*, 331(6023):1447–1451, 2011

**Examples**

- When you notice a fellow is frustrated because of their progress, you can acknowledge their experience and feeling. Then, you can tell them that struggling is natural when our brain learns new material. You can also share difficulties you have encountered in your research; for instance, when you had to change the initial plan because of some challenges or unexpected situations. You can tell them that even if their plans do not work out, they can document their observations, which will valuable for future researchers. Besides helping fellows re-frame, you can also also communicates that learning through experiments and mistakes is part of the creative process of science. [30]

- Be honest when providing feedback to a struggling fellow. You can affirm their talent, skill, and resilience. At the same time, you should be candid about the hard work they will need to do, perhaps in their project or courses they will need to take[31]. You can help students build their self-efficacy by being honest and provide constructive and specific feedback.

[30] See the paper Bell, R. et al. (2003) about how to help students become aware of the nature of scientific inquiry.

Randy L. Bell, Lesley M. Blair, Barbara A. Crawford, and Norman G. Lederman. Just do it? impact of a science apprenticeship program on high school students' understandings of the nature of science and scientific inquiry. *J Res Sci Teach*, 40(5):487–509, 2003

[31] Gail Horowitz. *Teaching STEM to first generation college students : a guidebook for faculty & future faculty*. Charlotte, NC : Information Age Publishing, Inc., 2019

*Fostering a growth mindset Mind* A mindset is the established attitude held by someone. People with a *fixed mindset* tend to think intelligence is fixed, and therefore they avoid challenges, give up when they encounter obstacles, see efforts as worthless, ignore constructive feedback and feel threatened by the success of others. On the other hand, people with a *growth mindset* believe that one's intelligence can be developed. They embrace challenges, sees efforts as paths to mastery, persists in setbacks, and inspired by other people's success, and therefore they are more likely to persevere and succeed[32]. We can model a growth mindset in our day-to-day interaction with fellows to foster a growth mindset.

[32] Carol S. Dweck. *Mindset: The New Psychology of Success.* Random House, 2006

**Examples** [33]

[33] These examples are from the website: Stanford Center for Teaching and Learning. Growth mindset. https://ctl.stanford.edu/growth-mindset

- Don't be afraid to say you don't know something; instead, tell fellows how you plan to find the answer. Demonstrating your learning process shows that knowledge is not innate, but can be obtained by conscious effort. You are also helping fellows learn how to learn, which helps them become self-directed learners.
- Ask open-ended and authentic questions so fellows can focus on the thinking process instead of obtaining a right or wrong answer. You can guide them to assess different approaches, and emphasize the reasoning.
- Avoid words that suggest a fixed mindset (e.g., do not talk about a "math" person) and use expressions that are indicative of a growth mindset (e.g., add "yet" at the end of a sentence that describes something that we do not master at this point in time).

**Activity** What are your takeaways from these strategies? List one or two things that you will try to help fellows feel that they belong.

## 1.5 *Creating an inclusive learning environment*

According to the Survey of Earned Doctorates, 593 female, compared to 1438 male students obtained doctoral degree in mathematics and statistics in a US institution in 2020. Among the 928 US citizens, only 30 are black or African American, 56 Hispanic or Latino, as compared to 118 Asian and 661 White (9 unreported). This statistics shows that female and minority students are underrepresented among doctorates compared to white or male students. While there are many reasons for this, it is important to realize that students from an underrepresented group face additional challenges in higher education because of stereotype, identity interference or previous negative experience. In this section, I focus on these additional barriers, and what you can do to help underrepresented students feel that they belong. Here, the term *underrepresented group* of students refer not only to students from a gender or racial minority group, but more generally to groups underrepresented in science, such as those from LGBTQ2S+ community, students with disabilities, and first-generation, low-income students.

### 1.5.1 Challenges to belonging in the academic setting

**Activity**

1. List three challenges you think underrepresented students might face in higher education.

2. Watch one or two of the following videos where students discuss challenges they face as members of underrepresented minorities in undergraduate institutions and reflect again on the challenges they face. What in the video surprised you?

   - Who We Are: First-Generation College Students Speak Out FLI at Stanford: the voices of first-gen and low income students[1]
   - "Ask us": What LGBTQ students want their professors to know[2]
   - What LGBT Students Want You To Know[3]

[1] *First-gen, low income and claiming a community: How FLI students are transforming the university for everyone*, Stanford magazine, Video URL: https://youtu.be/RRExHrSESxo
[2] Video URL: https://youtu.be/RRExHrSESxo
[3] Video URL: https://youtu.be/B-G4vA6TsX4

You may notice a few themes emerging from these videos. Underrepresented students may see few people of their identity at the university; First-generation students may feel academically unprepared and at a loss as to how to succeed in higher education (in our case, the data science workforce). Below, I summarize a few challenges underrepresented students might face.

*Identity interference* Because of unequal representation of different identities in a field of study, people might hold stereotypes of professionals in that discipline. For instance, one might think of a data scientist as white male wearing t-shirts and jeans, who spends hours in front of their computer writing code, from their impression after watching the movie "The Social Network". A student who does not conform to the stereotypical image might feel that their identity is incompatible with that of a data scientist. They may even feel that their identities *interfere* with being a successful data scientist. The clashing between identities, where one identity interferes another, is sometimes referred to as *identity threat*[34]. Research has shown that identity threat causes lower satisfaction and higher depression among female scientists[35]. While we typically focus on social identities such as race and gender,[36] it might be helpful to consider other aspects of one's life circumstances as well, such as being a mother, provider or a caretaker.

Finally, even without identity interference, students from under-

[34] Michael L. Slepian and Drew S. Jacoby-Senghor. Identity threats in everyday life: Distinguishing belonging from inclusion. *J Pers Soc Psychol*, 12(3): 392–406, 2021

[35] Isis H. Settles. When multiple identities interfere: The role of identity centrality. *Pers Soc Psychol Bull*, 30: 487–500, 2004

[36] Social identity typically include race, gender, socioeconomic status, sexual orientation, disabilities, religion and intersectionality, see

Northwestern Searle Center for Advancing Teaching and Learning . Social identities. https://www.northwestern.edu/searle/initiatives/diversity-equity-inclusion/social-identities.html

represented groups may feel they don't belong simply because there is a fewer of them in a discipline[37].

*Stereotype threat* *Stereotype threat* is the predicament that people feel that they are at risk of conforming to a stereotype about their social group. How does stereotype threat affect underrepresented students' performance? Think about a female taking a math test. She might think that if she does not do well in the test, then it confirms the stereotype that women cannot do well in math. Instead of focusing on the test, she has to divert her mental efforts to suppress these negative thoughts. The anxiety and distraction prevents her from performing optimally. In a seminal study, Steele and Aronson found that asking African American college students to indicate their race at the front of a test booklet caused their grades to drop[38]. Since this study, stereotype threat has been observed in many social groups and settings[39]. We note that stereotype threat is situational: for instance, while a female student might experience stereotype threat when taking a math test, the effect might disappear when the exam explicitly states that it does not measure their mathematics ability.

*Previous negative experiences* Underrepresented students are more prone to harrassment in an academic environment, including racism and discrimination. They are also more likely to experience microaggression in their daily life. [40] Derald W. Sue defines microaggression as "The everyday slights, indignities, put downs and insults that people of color, women, LGBT populations or those who are marginalized experiences in their day-to-day interactions with people." Microaggression can be intentional or unintentional; it can be verbal (e.g., "Everyone can succeed in this society, if they work hard enough." [41] ) or non-verbal (e.g., mistaking female doctor as nurse). It creates a hostile environment and may affect the physical health of those affected[42].

### 1.5.2 *How to create an inclusive learning environment*

By reading the challenges underrepresented students face in academic setting, you have accomplished the first step to creating an inclusive environment — by becoming aware. I now describe two strategies you can use to empower your students and creating a welcoming environment for everyone.

*Incorporating work in different application areas* Research shows that first-generation college students tend to be attracted to careers that are prosocial, i.e., has the potential to benefit their community or im-prove the well-being of others[43]. One way to help students feel they belong is to affirm their values and to demonstrate lead to prosocial outcomes.

[37] Catherine Good, Aneeta Rattan, and Carol S. Dweck. Why do women opt out? sense of belonging and women's representation in mathematics. *J Pers Soc Psychol*, 102:700–717, 4 2012

[38] Claude M Steele and Joshua Aronson. Attitudes and social cognition stereotype threat and the intellectual test performance of african americans. *J Pers Soc Psychol*, 69:797–811, 1995

[39] Stereotype threat. https://philosophy.rutgers.edu/climate-v2/climate-issues-in-academic-philosophy/stereotype-threat#datathreat. Department of Philosophy, Rutgers University

[40] See the article Campus Climate Survey reflects significant issues, Faculty Senate report details at the URL https://news.stanford.edu/report/?p=6010.

[41] In this example, the speaker attributes disparate outcome to people's laziness and denies the effect of systemic racism, see
Microaggressions part i: What are they, how are they harmful and what to do if you commit one? https://vpfo.ubc.ca/2021/03/microaggressions-part-1/, howpublished = Vice-President Finance and Operations Portfolio, University of British Columbia

[42] Jenée Desmond-Harris. What exactly is a microaggression? https://www.vox.com/2015/2/16/8031073/what-are-microaggressions

[43] Matthew C. Jackson, Gino Galvez, Isidro Landa, Paul Buonora, and Dustin B. Thoman. Science that matters: The importance of a cultural connection in underrepresented students' science pursuit. *CBE Life Sci. Educ.*, 15, 9 2016

**Examples**

- Try to learn about the value and interest of your students, and provide examples that aligns with their interest. For example, if students are interested in global warming, then you can use an animated visualization of the global temperature as an example of data visualization.

- Feature work of a diverse group of researchers. students are more likely to feel that they belong and persevere through challenges if they see people of their identity succeed. [44]

[44] You can even consider a spotlight assignment and ask students to share stories of a scientist who inspires them. This website(URL: `https://scientistspotlights.org/about-us/`) highlights scientists from a variety of background.

*Use micro-affirmations* The opposite of microaggression is *micro-affirmation*, which focuses on active listening, recognizing and validating experiences and affirming emotional reactions[45].

[45]

**Examples**

- Give heed to circumstances of each fellow. For example, if one of your students has to work in the afternoon, you can take the initiative to ask fellows to coordinate time that can work for everyone.

- Affirm fellows' contributions and celebrate their success. For example, when they asked a thoughtful question or make a stimulating observation, you can say "That's a very interesting observation!".

- You can ask for fellows' feedback and identify suggestions that you can implement.

**Activity** What can you do as a mentor to create an inclusive learning environment?

## 1.6 *Other topics in mentoring*

### 1.6.1 *Cross-racial mentoring*

Since DSSG fellows usually come from diverse cultural backgrounds, I want to discuss some challenges of cross-racial or cross-cultural mentoring and what you can do to address them. Materials in this section are adapted from Chapter 13 of *On being a mentor*[46].

[46] W. Brad Johnson. *On being a mentor: A guide for higher education faculty.* Lawrence Erlbaum Associates Publishers., 2007

*Challenges in cross racial mentoring* Mentees may feel stress and isolation if they have experienced bias, rejection or hostility before. They may worry that they need to discard their racial or cultural identity to fit in. They may feel distrust or skeptical about the support they would get from their mentors if they have experienced stereotypes, e.g., the stereotype that Asian students do not need mentoring. It might take some time for them to be familiar with how to be a successful mentee in the Western culture. For instance, because Asian culture focuses on collectivisim, Asian students might not want to overburden their mentor, and therefore they may find it difficult to communicate personal needs.

On the other hand, mentors might have misconceptions about working with students from racial groups different from their own. For instance, they may think that only same race mentoring can be effective. Or they may think there's no difference when mentoring students of the same or different race. They might even hold negative stereotypes; this might lead them to withhold their support for students until they have proved against the stereotype, or they might overcompensate or become overprotective.

To be an effective mentor, you should recognize the stereotypes you might hold, appreciate differences, and deliberately work towards enhancing your capacity to work with students from different cultures.

*Strategies for mentoring students from another race or culture*

- Recognize stereotypes, both those that are common and whose that you might personally hold.
- Appreciate individual differences. Rather than seeing students from different culture as a group, see each individual student as unique and complex.
- Work to enhance your cross-culture competence. For instance, spend time to learn the cultural heritage of your minority mentee, recognize racially related stress, and openly address cultural differences.
- Establish trust with your mentee.
- Listen for your mentee's preference when addressing racial differences.
- Encourage mentees to build secondary relationships, e.g., encourage African American students to connect with African American student association.

# 2 Development of your project

As a technical mentor, you may need to take on the role of a project manager, carefully planning the project, guiding your team along the way, and facilitating communication with your project partner. In this section, I focus on three topics:

- *Collaborating with your project partner* DSSG team often works with an organization (we refer to them as the "project partner") to solve a problem or answer a question. This section discusses strategies about how you can *initiate* and *sustain* collaboration with your project partner to ensure that DSSG project successfully addresses their need. You will outline a project milestone at the end of this section.
- *Leading effective team meetings* During DSSG, you will meet with your team daily to brainstorm ideas, share progress, provide feedback, and set goals. We describe several discussion moves to help you achieve this during team meeting, while helping fellows grow as data scientists and team players.
- *Housekeeping* This section offers a list of questions to help you decide how to structure and organize team work and project materials.

## 2.1 Collaborating with your project partner

As a scholar, you might be more familiar with curiosity-driven research, where you determine a topic of your interest and leave implementing your theory or method to practitioners, as compared to *collaboration* between a research team and a project partner. Though it takes some effort to engage your project partner, your endeavor will be fruitful because DSSG projects have the potential to make a real impact, not only through innovative research, but also through addressing an issue that affects people!

I start with a case study (from my personal experience): what elements in this example enhance collaboration and how can this project be improved?

### 2.1.1 *A case study*

**Activity** Read the description of the following DSSG project. What did the team do to facilitate collaboration with the project partner? How can the project be improved?

In the project "Forecasting Platelets Blood Bag Demand to Reduce Inventory Wastage at the Stanford Blood Center", we collaborated with a medical officer at Stanford Blood Center to develop an inventory management strategy for platelets blood bags.

At the start of the program, our project partner gave us a tour around the blood center. We saw platelets blood bags in the refrigerator, maintained at a stable temperature and being gently shaken from side to side. After the tour, he told us what platelets are and the challenge in managing the platelets inventory. Platelets are cell fragments in our blood that form clogs and stop or prevent bleeding. A patient usually requires platelets transfusions in one of two situations: when they experience severe blood loss during a surgery, or when they do not have the capability to produce platelets. The blood center orders a certain number of blood bags according to whether it is a weekday or weekend. Because platelets blood bags have a short shelf-life, it is easy to incur waste by ordering too large amount. [1] At the same time, the blood center wants to prevent shortage because of the critical importance of platelets. Therefore, they want to develop a data-driven method to determine how many blood bags to order every day.

[1] Platelets is particularly valuable because a donor can only donate platelets twice a month

During our discussion with our project partner, we noticed that they were concerned about using a "black-box" strategy, where an algorithm recommends a number without any explanation. Therefore, we decided to predict platelets demand first, leaving to others the task of leveraging our predictions to devise an inventory management strategy. After exploratory data analysis, we noticed that sometimes a patient needs several blood bags during a serious surgery, while other times, one patient comes to the hospital regularly for platelets transfusion. Based on this observation, we decided to separately consider patients who undergo severe surgery and those who need regular transfusions. We hypothesized that separating the two sources of usage would enable us to arrive at a more accurate prediction.

We implemented different models to predict platelets demand and evaluated our models using past data. We also met with our project partner each week to present our progress and obtain his feedback on our approaches. We applied our model to historic data and found that it was able to predict platelets demand more accurately compared to predicting by the average number of blood bags used, which does not use any information about demand. However, our project partner was not able to deploy our model because the hospital changed their data base and we did not discuss how to implement our model at the end of the program.

### 2.1.2 *Strategies for effective collaboration*

The DSSG project in the case study was a collaborative effort in the sense that

1. The project partner brought a problem, data, domain expertise and their value orientation.
2. The DSSG team brought analytical skills, such as data visualization and modeling. The project partner's domain knowledge inspired the team's modeling approach.
3. The DSSG team and the project partner decided project objectives together.
4. Both parties were aware of their roles in the project and dedicated a fixed time to discuss progress and provide feedback.

This kind of collaboration bears certain resemblances to community-based participatory research (CBPR), which can defined as "an approach to research that involves collective, reflective and systematic inquiry in which researchers and community stakeholders engage as equal partners in all steps of the research process with the goals of educating, improving practice or bringing about social change." [2] I draw some ideas from studies of CBPR, mainly from the following two articles, and complement them with examples of DSSG projects or CBPR.

- *Community-Engaged Research with Community-Based Organizations: A Resource Manual for UCSF Researchers* by Pasick, R. et al. [3]
- *Meaningful community collaboration in research* by Swann S.A., Campbell, A.R. and Murray, A.R.C.V.J.N.M.C. [4]

*Engage your project partner in every stage of your project* You should arrive to a shared problem definition, analysis strategy, interpretation of results [5]. You should take the initiative to organize regular meetings and communicate questions, progress and concerns.

[2] Marie-Claude Tremblay, Debbie H. Martin, Alex M. McComber, Amelia McGregor, and Ann C. Macaulay. Understanding community-based participatory research through a social movement framework: a case study of the kahnawake schools diabetes prevention project. *BMC Public Health*, 18(1):487, 2018

[3] Rena Pasick, Geraldine Oliva, Ellen Goldstein, and Tung Nguyen. *Community-Engaged Research with Community-Based Organizations: A Resource Manual for Researchers*. Community Engagement Program, Clinical and Translational Science Institute at the University of California, San Francisco

[4] Shayda A. Swann and Amber R. Campbelland Valerie J. Nicholson Melanie C.M. Murray. Meaningful community collaboration in research. *B. C. Med. J.*, 62(9):340–341, 2020

[5] Lisa Cacari-Stone, Nina Wallerstein, Analilia P. Garcia, and Meredith Minkler. The promise of community-based participatory research for health equity: A conceptual model for bridging evidence with policy. *Am J Public Health*, 104:1615–1623, 2014

**Example: DSSG project "Algorithmic Equity Toolkit" [6]**

Community civil activists have been concerned about the expansion of automated decision system (ADS), such as facial recognition systems. Because such systems are hard to comprehend, it is challenging to determine their impact. The DSSG team developed a toolkit, which includes education material about the technology, a guide to help community civil activists distinguish an ADS from a surveillance tool, and a questionnaire to identify potential harms of such tools when deployed by government agencies. In the project website, fellows described their process of developing their toolkit. The team met with the project partners to understand their work, mission and need. [7] The team then narrowed down the audience of their toolkit to focus on civil rights advocacy, grassroots organizations, and members of the general public [8]. Each week, the team designed a new prototype, solicited feedback from the project partner, and updated their design based on the feedback. To evaluate their toolkit, they designed a survey to get both objective data, e.g., how well their project partner understand relevant concepts, and subjective feedback, e.g. how useful is the toolkit. Finally, they invited data scientists to evaluate the accuracy of their education material.

[6] Corinne Bintz, Aaron Tam, Vivian Guetler, and Daniella Raz. Algorithmic equity toolkit. https://uwescience.github.io/AEKit-website/, 2019. Data Science for Social Good, University of Washington

[7] The project partners in this work are the American Civil Liberties Union, the members of the Community Centered Tech Coalition, specifically Densho and the Council on American Islamic Relations.

[8] Michael Katell, Meg Young, Dharma Dailey, Bernease Herman, Vivian Guetler, Aaron Tam, Corinne Bintz, Daniella Raz, and P. M Krafft. Toward situated interventions for algorithmic equity: Lessons from the field. pages 45–55. Association for Computing Machinery, 2020

*Promote co-learning* You and your project partners might view a problem from different perspectives, and you can both learn from each other. For example, the DSSG team can gain domain knowledge while the project partner can learn about the technical approach.

**Example: DSSG project "Developing Ensemble Methods for Initial Districting Plan Evaluation" [9]**

In this project, fellows developed a user guide on how to generate district maps that account for state-level redistricting rules, how to evaluate trade-offs in selecting maps, and how to test whether existing maps are fair and representative. In a blog post [10], fellows described how the project partner influenced their conceptual understanding of the project. One fellow mentioned "Engaging with stakeholders has helped ground our work in reality. By talking with experts from a variety of backgrounds, [...], we have gained insight on the practical considerations involved with redistricting." Another fellow mentioned "Meeting with people who are engaged in redistricting has given us invaluable insights into this process and helped us to develop outputs with these stakeholders in mind." Learning from their stakeholders motivated the team to orient their work towards practical application, bridging the gap between theory and practice that often occurs in research.

[9] Rowana Ahmed, Katherine Chang, Ryan Goehrung, and Michael Souffrant. Developing ensemble methods for initial districting plan evaluation. https://uwescience.github.io/DSSG2021-redistricting-website/, 2021

[10] Emily Keller. Data science for social good team builds tools to support fairness in computational redistricting, 2021. URL https://escience.washington.edu/data-science-for-social-good-computational-redistr eScience Institute, University of Washington

*Be flexible* Many challenges can arise during the project. For example, your project partner may not be able to obtain the data on time, or they may not be available for a certain period of time. You will need to be flexible during unforeseeable circumstances and be open to taking alternative approaches.

**Example: DSSG project "Geography, Equity, and the Seattle $15 Minimum Wage Ordinance"** [11]

This project aimed to understand the effect of increasing minimum wage on low-income community. Particularly, the team focused on three issues pertinent to the lives of residents: housing burden, residential displacement and commute. The team used Washington Merged Longitudinal Administrative Data (WMLAD) for their analysis. Because this data contains highly sensitive information, such as employment, earning, residential address, etc., it is only accessible through a remote enclave with no internet. To use this data, the fellows developed a specialized workflow and met with state IRB members to discuss privacy considerations for using this data. Instead of becoming discouraged because of the inconvenience, the team reaffirmed their commitment to privacy and equity, and embraced the challenge as an opportunity to learn how to write efficient code. They also created a workflow and recommendations for future researchers who work under the same constraint.

[11] James Lamar Foster, Delaney Glass, Christopher Salazar, and Mahader Tamene. Geography, equity, and the seattle $15 minimum wage ordinance. `https://uwescience.github.io/MinWA/`, 2021

*Be respectful and demonstrate curiosity* Respect and curiosity are important to build trust in any relationship; it is especially important in a DSSG project since the DSSG team is an outsider to the community.

**Example: the "Messengers for Health" project** [12]

This project is a collaborative effort between an American Indian community and academic researchers to increase Crow women's participation and knowledge regarding cervical cancer screening and prevention. The researchers intentionally built trust between community and themselves through many efforts. They attended anti-bias training to gain awareness of their personal history and bias. They learned about the history and the culture of the tribe, as well as the historical context of their research. They attended social and cultural events. They were transparent about their intentions, and they acknowledged and engaged the expertise of the community partner. Their efforts resulted in an 11-year long partnership. Over the years, more community members became willing to participate in the project. The research team were also approached by male community members to develop a project on men's health issues on the reservation. The Messengers for Health project were also invited to Crow Fair, a significant social event for the tribe.

[12] Suzanne Christopher, Vanessa Watts, Alma Knows His Gun McCormick, and Sara Young. Building and maintaining trust in a community-based participatory research partnership. *Am. J. Public Health*, pages 1398–1406, 2008

*Be clear about your expectations* For instance, be upfront about how often you expect to meet with your project partner, what decision they need to make and when do you expect their response. You should be transparent about your responsibilities and how much time you and the team will devote to the project. If you plan to write an academic paper, you should specify the roles of the DSSG team and the project partners, and if possible, provide guidelines that both of you will follow.

*Be committed to disseminating and transferring knowledge* Not only should you provide reproducible and documented code, but also you should report your analysis and communicate your results in a way that people without data science expertise can understand them.

**Example: "Working with the Los Angeles House and Ball Communities"** [13]

Researchers in this study aimed to develop interventions to reduces AIDS risk among the House and Ball communities in Los Angeles. Houses are primarily composed of African American men who have sex with men under 30 who perform at Balls, which are competitions focused on dance, gender expression, sexuality and fashion. The researchers reflected that designing intervention is difficult, because "the community members need to think through what is it about their communities that may put their members at risk", while at the same time, "it can also be an experience that empowers the community to not only work to address those aspects of risk but also identify and embrace the sources of support that exist and build on those sources to create a safe and healthy environment. "

[13] Southern California Clinical and Translational Science Institute Community Engagement. Toolkit for developing community partnership. https://sc-ctsi.org/uploads/resources/DevelopingCommunityPartnerships_Toolkit.pdf

I hope you find inspirations from these case studies and feel more confident about collaborating with your project partner! To conclude, I highlight two elements underlying these suggestions: *mindset* and *structure*. You bring to the partnership not only your research expertise, but also your creativity, patience, humility, respect, and commitment to addressing issues faced by the community [14]. To facilitate communications, you should set up regular meetings and clarify how to communicate with your project partner. The goal of first few meetings is for you and your project partner to come to a shared vision for the project.

[14] Rena Pasick, Geraldine Oliva, Ellen Goldstein, and Tung Nguyen. *Community-Engaged Research with Community-Based Organizations: A Resource Manual for Researchers*. Community Engagement Program, Clinical and Translational Science Institute at the University of California, San Francisco

### 2.1.3 *Deciding on a project vision with your project partner*

A project vision provides focus and inspiration at the beginning, and it will continue to evolve as the project progresses. Below, I provide a few prompts you can discuss with the project partner.

1. *What questions do you want to answer? What problems do you want to address?*

   Often times your partner have a broad goal in mind, and you will need to work with them to refine it to a list of concrete and tangible research questions. For instance, in the DSSG project "I-405 HOT Lane Equity" [15], the team used Washington State Department of Transportation traffic data, combined with other data sources, to identify whether constituents would be unduly impacted by highway tolls and other transportation management policies. The team narrowed down their project scope into three specific questions: how are the I-405 HOT lane facility's benefits and costs distributed among and within different groups of users? Are there any inequitable distributions? What policy implications or recommendations can we make to address potential inequities?

   As another example, in one CBPR study, researchers originally planned to identify options to prevent cervical cancer in an under-resourced community in Cape Town, South Africa. After discussing with community members, they refocused their research interest from cervical cancer to "cervical health," acknowledging the broad epidemiological and social impact on women's health [16].

2. *What is the impact of answering this question/addressing this problem?*

   Understanding the benefit of the project directs your team to focus on the need of your project partner and provides an important perspective from which you evaluate your work. It also helps to motivate you and your team as you work through many challenges during the program.

3. *Which community is impacted by the project?*

   I stress the community as a stakeholder, even though you do not directly engage with them, because they are often directly impacted by your research. Bearing them in mind motivates you to confront ethical implications of your analysis. During the project, you should continuously ask questions such as "from whom are the data collected?", "what source of bias is encoded in the data?", "how to ensure our product is fair?" etc.. DSSG programs emphasize ethical problem solving and the starting point is recognizing the community impacted by your research.

   Neglecting bias in the data can lead to grave consequences. For example, the city of Los Angeles employed a predictive policing company to decide which neighborhoods in LA to patrol more heavily [17]. The model was based on both historical crime data as well as crime pattern in recent periods. Because historical data encodes past bias in the criminal justice system, inferring current

[15] Shirley Leung, Cory McCartan, Kiana Roshan Zamir, and CJ Robinson. I-405 hot lane equity. https://uwescience.github.io/hot-lane-equity/, 2019

[16] Maghboeba Mosavel, Christian Simon, Debbie van Stade, and Mara Buchbinder. Community-based participatory research (cbpr) in south africa: Engaging multiple constituents to shape the research question. *Soc. Sci. Med.*, 61 (12):2577–2587, 2005

[17] Los Angeles Police Department ended their contract with the company, PredPol, in 2020. See

Johana Bhuiyan. LAPD ended predictive policing programs amid public outcry. a new effort shares many of their flaws, 2021. The Guardian criminal activity based on past data tends to resulted in disproportionate patrolling of neighborhoods of color [18,19]. In this situation, keeping the attention on the community impacted — the residents of LA, and in particular the marginalized groups — would lead to a different approach. For instance, one might question the historical bias in the data, consider an auditing process for an algorithm, or even think about how to build trust from community.

[18] Catherine D'Ignazio and Lauren F. Klein. *Data Feminism*. The MIT Press, 2020
[19] Stop LAPD Spying Coalition. Before the bullet hits the body – dismantling predictive policing in Los Angeles, 2018

4. *What deliverable would be useful for you? What aspect of the deliverable is important to you?*

   There are many types of deliverables you can provide, for example, detailed analysis, an algorithm or a software, a policy recommendation etc.. Clarifying deliverables can help you prioritize your work. Your project partner's value can also shape your choice. For example, when the project goal is prediction, type II error might be more important than type I error, as in the case in our partnership with the blood center. Your project partner might value the transparency and explainability of your algorithm. Sometimes, they most appreciate that your results are communicated in clear and nontechnical language.

5. *What are some resources for the project? What data are available?*

   You can consult your project partner about background information and project context, e.g., related literature, news stories and legislation. You might be able to build your work on previous research the project partner has conducted on a similar problem.

   Since data is critical to DSSG, you need to obtain the data source before the program starts so the project can hit the ground running. If your project partner has collected the data, you need to examine them critically: are they sufficient to address the question? If the project partner still needs to collect more data, then you can provide feedback on what data they can collect.

6. *When should we meet and what is your preferred way to communicate?*

   You can suggest a few options to your project partner, e.g., email, slack, in-person meeting or Google meet. You should also clarify who you can contact for specific questions. You should schedule meetings before the program starts and clarify with your project partner when will be unavailable.

### *2.1.4 Project milestones*

After a few meetings, you should have some ideas about what the project can accomplish, who will participate and with which roles, and what resources are available. Now it is time to draft milestones!

The milestones makes it easy to communicate progress with your project partner, and provide a reference with which you can troubleshoot. I will borrow ideas from the book *10 Steps to Successful Project Management* [20]. Since everyone has a different style, I encourage you to explore approaches that are most suitable for you.

[20] Lou Russell. *10 Steps to Successful Project Management*. Association for Talent Development, 2007

*Step 1* Outline the goals of the DSSG project.

*Step 2* List what the project team needs to do in order to achieve the goals at a high-level, such as "fit a statistical model".

*Step 3* Identify what and when you need input or feedback from your project partner.

*Step 4* Estimate how long each task would take. At this point, you might break down each task into more granular activities. For example, you can write detailed steps for tasks you listed in step 2, such as "fitting a random forest", "select parameters" and so on.

*Step 5* Write detailed steps in a document. You can highlight project milestones in the document.

*Step 6* Visualize the project trajectory and dependencies using a Gantt chart (Figure 2.1 shows an example). [21] Each row in the $y$-axis indicates tasks one individual should accomplish, but you can also devote one row to each task. The $x$-axis indicates time or date. The starting and ending point of the project is marked by diamond shapes, and the duration of each task is marked by a rectangle. The relations between tasks determine the starting time of different tasks, but you can also explicitly note task dependencies by connecting tasks along a trajectory.

[21] Gantt chart was named after Henry Gantt, and was designed around 1910-19

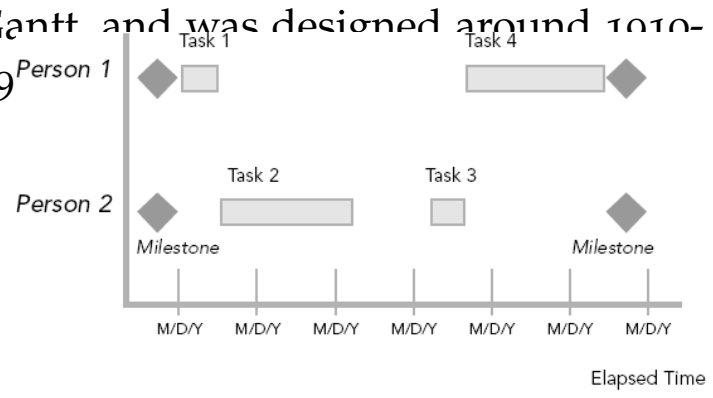

Figure 2.1: Example of Gantt chart (Figure 6.3 in the book)

| |
|---|
| **Activity** Draft the project milestones and discuss them with your project partner. |

## 2.2 *Leading effective team meetings*

An effective team meeting ensures everyone is on the same page and sets a positive tone for the day. During the meeting, the team can share progress, brainstorm ideas and divide tasks. Unlike a workplace setting, where a project manager would assign tasks to each team member, you will achieve these functions by facilitating team discussions. Through discussions, the fellows can discover a solution together, practice critical thinking and reflect on their work. [22]

In this section, I will discuss the structure of a discussion and what language you can use at each step. I also describe two common scenarios you might encounter during team discussions.

[22] In the article *The dreaded discussion: Ten ways to start*, the author stresses the importance of letting students discover themselves: "The key to effective retention of learning, I believe, is in owning the discovery. Emerson wrote in his journals that a wise person 'must feel and teach that the best wisdom cannot be communicated (but) must be acquired by every soul for itself'. My primary strategy as a teacher is to structure situations in which students have as many opportunities as possible to acquire wisdom for themselves; that is, to own the discovery of a new learning insight or connection and to express that discovery to others".

Peter Frederick. The dreaded discussion: Ten ways to start. *Improving College and University Teaching*, 29(3): 109–114, 1981

### 2.2.1 Elements of a discussion

The three basic components of a discussion are *questioning*, *listening* and *responding*. By asking questions, you can guide the discussion, clarify information and probe fellows' learning. By listening to fellows' responses, you can gauge their understanding or misconceptions. Responding allows you to follow-up with more questions or information.

*Posing a question* Different questions elicit different types of answers and serve different purposes. The following are a few types of questions you can ask, and you can find more types of questions in the handout *A typology of questions*[23].

[23] The Derek Bok Center for Teaching and Learning, Harvard University. Leading discussions. https://bokcenter.harvard.edu/leading-discussions

- *Clarifying question* Ask fellows to explain their ideas or work.

  **Examples**

  - What's an example of what you said? What do you mean that ...?
  - I think you meant ..., is that correct?
  - I am confused about ..., can you explain more?

- *Challenging questions* Ask for supporting evidence, or ask fellows to explain their reasoning or assumptions.

  **Examples**

  - How do you know this?
  - What information do we need to verify this hypothesis?

- *Relational questions* Encourage fellows to connect different pieces of information.

  **Examples**

  - How does this new information relate to our work?
  - How do these two ideas compare?

- *Open-ended questions* Encourage fellows to brainstorm new ideas.

  **Examples**

  - Any ideas on what features we might consider in our model?
  - What should we include in our deliverable?

- *Priority questions* Encourage fellows to synthesize what has been discussed to a list of actionable items.

  **Examples**

- How would you go about selecting variables?
- What's the first step?
- What will you do next?

*Listening to responses* Listen actively and observe the quality of fellows' responses: do you notice any misconceptions? are they able to elaborate on their ideas? are their thinking logical? how do they interact with each other? You might want to wait for a few seconds before you respond to encourage fellows to reply to each other.

*Follow-up* There are three types of follow-up moves:

- *Confirming* You can confirm a fellow's response, or draw connection between their idea with other research. For example, you can say "That's a good idea! I'm glad you brought this up!", or you can explicitly tell them that you agree.
- *Rejecting/Challenging* You might point out misconceptions or inconsistencies in a fellow's response. You can also challenge them to clarify or extend their ideas. For example, you might ask "Can you apply this idea to another situation, if so, what else do you need?".
- *Eliciting responses* You can encourage other fellows to reply. For example, you can ask "What do you think about what we just said?".

You can read more about follow-up moves in the handout *Techniques for responding to students in discussions* [24] and Chapter 3 in the book *The art of discussion-based teaching: opening up conversation in the classroom*[25].

[24] The Derek Bok Center for Teaching and Learning, Harvard University. Leading discussions. https://bokcenter.harvard.edu/leading-discussions

[25] John E. Henning. *The art of discussion-based teaching : opening up conversation in the classroom.* New York : Routledge, 2008

### 2.2.2 *Structure of a discussion*

Now that you have seen basic elements of a discussion, you can use them throughout a meeting. Yet you might still be wondering: How do I start a discussion? What do I do when discussion goes off track? In this section, I discuss five discussion moves: framing the discussion, encouraging deeper answers, broaden perspectives, refocusing the discussion, closing discussion. Materials here are based on the presentation *The art of leading discussions for sustainable learning* [26].

[26] Andrea Taylor. The art of leading discussions for sustainable learning. https://drive.google.com/file/d/1kpmf36PwnZS1MHwaPATMRjgixfQ6GU0Y/view

*Framing a discussion* At the opening, you should let fellows know what topic do you want to discuss and how is it related to their learning or the project; this is the guiding theme that you will return to throughout the discussion. Perhaps you want to clarify a concept together, to solve a problem, or to solicit opinions. It might be helpful to set norms with your team at the beginning as well. For example, you can remind them that they should be respectful towards each other and everyone should participate equally.

*Encouraging deeper answers* After you solicit a few responses, you can encourage fellows to think deeper. For example, you can ask them to explain, describe evidence, or draw connections with other ideas. Make sure to be curious when asking questions.

*Broaden perspectives* Now that you have discussed one or two ideas in deep, you can encourage fellows to think about alternative ideas. For example, you can say "This is a great perspective, let's hear another perspective", or "Are we missing anything? What issue have we not considered?"

*Refocus the discussion* Sometimes, you need to re-focus the discussion to one specific point. For instance, you can say "We have a few options on the table, let's focus on one point at a time.", or you can tell them you want to focus on one idea by saying "Let's hold off that thought for a moment and come back to that later. Let's discuss this point first". You can also ask fellows how they might reconcile different ideas.

*Closing the loop* At the end, you should summarize what has been discussed, or you can ask fellows to summarize . You might ask them "What are your takeaways?", "How does what we discussed today relate to the project?", or "Why should we care about these issues?"

### 2.2.3 *Challenges you might encounter*

I now describe two scenarios you might encounter during your team meetings. The suggestions are motivated by several sources [27,28,29,30].

> **Activity** Think about how you would address the issue before going through the strategies. If you are working with others, then I encourage one of you to lead a short discussion about each scenario and use it as an opportunity to try the discussion moves.

*Unequal participation* You notice that one or two fellows are particularly active and tend to dominate the discussion. At the same time, another fellow is left out.

**Strategies** If you notice that some fellows do not participate much, you can encourage them to speak up by making them feel more comfortable about sharing. For example,

[27] The Derek Bok Center for Teaching and Learning, Harvard University. Leading discussions. https://bokcenter.harvard.edu/leading-discussions

[28] Yale Poorvu Center for Teaching and Learning. Effective class discussions. https://poorvucenter.yale.edu/EffectiveClassDiscussions

[29] Santiago Mejia. Leading effective discussions. https://teaching.uchicago.edu/resources/teaching-strategies/leading-effective-discussions/. Chicago Center for Teaching and Learning

[30] Eberly Center Teaching Excellence & Educational Innovation, Carnegie Mellon University. Discussions. https://www.cmu.edu/teaching/designteach/design/instructionalstrategies/discussions.html

- You can start the team meeting with an icebreaker to help fellows get to know each other.
- You can invite fellows to participate by "warm calling": letting the fellow know that you would like them to answer a specific question and ask them to prepare beforehand.
- You can remind fellows of their strengths, that you value their perspectives and there is no right or wrong answers. You can remind them that this is a learning process for everyone on the team.

If a few fellows tend to dominate the discussion, you might want to set some norms with the team. For instance, they might agree not to interrupt, and make sure everyone has shared before speaking a second time. You can also set up a structure to ensure everyone can participate; for instance, you can ask fellows to talk in turns.

Though in this scenario you noticed one fellow did not participate, in practice you might not be aware that some voices are missing because you might be focused on leading the meeting. To avoid blind spots, you should keep track of fellows' participation. You can also meet with them individually to ask how they feel during team discussions, and usually they will bring up their frustration during individual conversations.

*Fellows not responding to each other* You notice that you are the only one responding to fellows, but they are not responding to each other.

**Strategies** Team meeting is a good place for fellows to learn how to work together, build on each other's ideas, and provide constructive feedback. If you notice that they do not respond to each other, you can remind them to do so because it is important to catch up with each other's work and decide together what's the best way for the team to move forward.

Sometimes fellows may not know how to respond in a discussion, and in this case you can demonstrate active listening and quality responses. You can also teach fellows different roles in a discussion[31].

Finally, you can explicitly ask for fellows' opinions. For example, you can ask one fellow to comment on the connection between another fellows' work and their own, or ask a question.

31 Stephen D. Brookfield and Stephen Preskill. *The Discussion Book : 50 Great Ways to Get People Talking*. San Francisco, CA : Jossey-Bass, a Wiley Brand, 1 st edition, 2016

## 2.3 *Housekeeping*

As the team leader, you should guide fellows to organize their project materials and help them collaborate. The following are some questions you can think about before the project starts. I suggest that you start with an idea of your own and also work with your fellows to agree upon something you would do as a team, and be consistent throughout the program.

1. How shall we communicate within the team? (e.g., when should we use slack and when to use email?)

2. How shall we organize the project folder? (e.g. you might want to create a few folders, such as resources, tasks, presentations and meeting notes. As the project progresses, the mean might need to create more folders to keep track of new notes)

3. How shall we organize the our code? (e.g. you can use *Cookiecutter data science*[32] as a starting point, but usually how you organize your code will depend on your project)

4. How shall we communicate progress and provide feedback to each other? (e.g. comment on each other's work during team meeting)

5. How would you keep track of the task of each team member? (e.g. perhaps everyone can write down their tasks in a shared document, or use task list in Github)

6. How shall we communicate with project partner/faculty member? (e.g. perhaps you want to prepare some slides each week, or you might create a notebook and each fellow can rotate to take notes)

**Activity** Think through these questions, and discuss with fellows during the first week.

[32] Cookiecutter data science. https://drivendata.github.io/cookiecutter-data-science/

# 3
# *Fellow's growth as data scientists*

For many fellows, the Data Science for Social Good program provides a valuable and engaging learning experience. Because the projects are interdisciplinary, fellows usually need to acquire specific skill; for example, they might need to learn game theory or use Structured Query Language (SQL). In addition to technical skills, they learn how to collaborate when working on a complex project, e.g., how to convey ideas, keep each other updated, share intermediate products, and come to a consensus. Besides interacting with each other, they might need to present their work to a community partner who usually does not have technical expertise, and thus fellows need to be able to explain concepts in plain language. Even more, DSSG may be the first time they apply their data science skills in an new scenario, where they identify an appropriate procedure, implement their ideas and refine their methods. Through this process, fellows deepen their understanding and generate new insights. All of these are new experiences, and in fact, they might be intimidating for some fellows. You should be prepared to communicate to your team that DSSG provides a strong learning experience and that you are committed to supporting them throughout the process.

The many possible areas of growth are mirrored in the breadth of approaches you can use to support them. For example, to help fellows gain technical skills, you can give them a group or individual tutorial, or you can probe their learning in conversations. To enhance communication skills, you can model effective communication and provide a list of suggestions to help fellows improve their presentation. To guide fellows to reflect on their learning, you can ask them to plan their approach, describe their reasoning, and provide each other feedback. These are but a few examples. In this section I will focus on three instruments to support fellows' learning: (1) cultivating an collaborative learning environment, (2) cultivating meta-cognitive skills, and (3) incorporating active learning in your tutorials. There are of course many other teaching strategies, and I pick these three because they are particularly applicable to the structure of the DSSG program. I refer you to the following two books if you are interested to learn more about teaching and learning.

- *How learning works: Seven Research-Based Principles for Smart Teaching* by Ambrose, S.A. et al. [1]
- *The ABCs of How We Learn: 26 Scientifically Proven Approaches, How They Work, and When to Use Them* by Schwartz, D.L., Tsang, J.M. and Blair, K.P. [2]

[1] Ambrose Susan A., Bridges Michael W., DiPietro Michele, Lovett Marsha C., and Norman Marie K. *How Learning Works: Seven Research-Based Principles for Smart Teaching*, volume 1 st of *The Jossey-Bass Higher and Adult Education Series*. Jossey-Bass, 2010

[2] D. L. Schwartz, J. M. Tsang, and K. P. Blair. *The ABCs of How We Learn: 26 Scientifically Proven Approaches, How They Work, and When to Use Them*. W W Norton & Co., 2016

Before diving into the three approaches, I discuss the cognitive process of learning. This background will help you understand why some teaching strategies are effective, as well as providing with a framework for you to evaluate teaching strategies that you might encounter in the future.

## 3.1 *What does learning really mean?*

**Activity** Throughout your academic journey, has the meaning of learning changed for you? What does it mean for you to *learn* a data science concept, such as cross-validation[1]?

[1] James, G., Witten, D., Hastie, T. & Tibshirani, R. (2013) *An Introduction to Statistical Learning: with Applications in R* Springer New York, NY.

You probably notice that learning happens at several levels. At the beginning, you learn what cross-validation *is*, then you learn how to *implement* it in a statistical software, and afterwards you may *apply* it in a project. You might *decide* that standard cross-validation does not apply because your data form a time series, so you searched and found one potential approach called "evaluation on a rolling forecasting origin" [3]. These different *cognitive processes* proceed from lower to more advanced levels. During graduate school, you may consider that you have understood a concept if you are able to *teach* it to someone else, or that you are able to connect it with other domains. These are the *human dimension* of learning.

[3] Rob Hyndman. Cross-validation for time series, 2016

### 3.1.1 *The cognitive process of learning*

In this section, I describe the cognitive process of learning through Bloom's taxonomy, which is a widely adopted framework to articulate educational goals in terms of cognitive activities. The original Bloom's taxonomy was published in *Taxonomy of Educational Objectives* (1956) [4]. We will instead focus on the revised taxonomy, published in the book *Taxonomy of Educational objectives: the classification of educational goals* (2021)[5].

[4] Patricia Armstrong. Bloom's taxonomy. `https://cft.vanderbilt.edu/guides-sub-pages/blooms-taxonomy/`, 2010. Center for Teaching, Vanderbult University

[5] *A taxonomy for learning, teaching, and assessing: a revision of Bloom's taxonomy of educational objectives*. Longman, New York, complete edition, 2001

The revised taxonomy categorizes cognitive activities of learning into six levels: remember, understand, implement, analyze, evaluate, and create (see Figure 3.2). I briefly describe each one below, and you can find more details from the book.

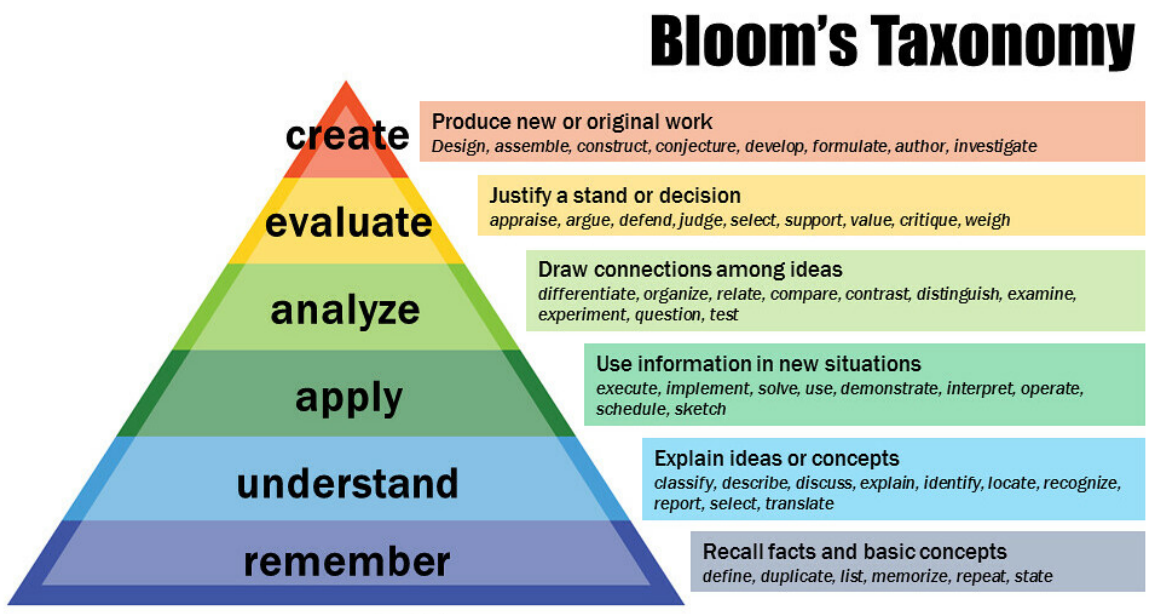

Figure 3.1: The revised Bloom's taxonomy is often represented as a pyramid, with the lower level process on the bottom and the higher level processes on the top.

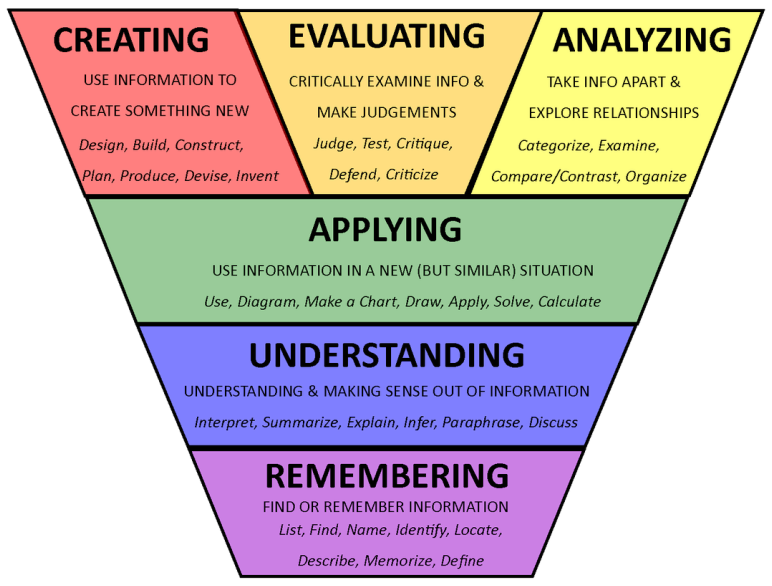

Figure 3.2: Higher level cognitive processes are sometimes represented without hierarchy.

- When you **remember**, you are able to retrieve information from long-term memory. The information can be factual, conceptual or procedural. For instance, you might remember the names of iconic impressionist paintings (factual); you may recall that the sample mean converges to a normal distribution as sample size increases (factual); or you might remember the steps to carry out a nonparametric bootstrap (procedural).

- When you **understand**, you are able to connect current knowledge with previous knowledge. For instance, you may notice that binomial distribution is an example of sum and thus also approaches a standard normal distribution when appropriately standardized. You can represent information in different forms; for instance, you can sing a tone when you see a note, or you can explain photosynthesis using both chemical equation and a diagram. You are able to provide examples or classify different instances; for example, you will can classify paintings into their corresponding art movements. You are also able to explain and troubleshoot.

- When you **apply**, you are able to execute a procedure by following previous examples, and you are able to choose a procedure when it is appropriate.

- To **analyze**, you are able to differentiate steps in a procedure or an argument. You can organize materials into a coherent structure, and you are able to distinguish the point of view underlying a line of reasoning from the supporting arguments.

- When you are able to **evaluate**, you can check logical inconsistencies and critique the effectiveness of a given method.

- When you **create**, you can represent a problem, generate a solution, lay out steps to implement your approach and execute the steps.

**Activity** Pick a data science concept that you want to teach DSSG fellows (you can choose a familiar topic if you do not not need to prepare a tutorial).

1. How can you assess if fellows are able to understand this concept? What about if they are able to evaluate a related procedure?
2. Write down three learning objectives. A learning objective is something fellows will be able to do by the end of your tutorial.[1] Your lesson can have multiple learning objectives, spanning multiple cognitive activities. For example, "fellows will able to construct a logistic regression model for binary response data, make inference for model parameters and predict odds at a given observation."

[1] You may also want to determine how well fellows know about a topic before your tutorial and design your lesson with that in mind.

### 3.1.2 *The holistic approach to learning*

Bloom's taxonomy focuses on the cognitive aspects of learning, whereas Fink's learning objective, formulated in the book *Creating Significant Learning Experiences: An Integrated Approach to Designing College Courses*[6], takes into account several other aspects. [7] Fink categorizes learning objectives into six groups; besides the cognitive processes described in Bloom's taxonomy (corresponding to the first two groups — foundational knowledge and application), it also points to students' development in other dimensions. These four categories are:

[6] L. Dee Fink. *Creating Significant Learning Experiences*. John Wiley & Sons, Incorporated, 2 nd edition, 2013

[7] Fink characterizes a "significant learning experience" as one where students are engaged, and has the potential to improve their lives in some important ways.

- *Learning how to learn*: students are able to (1) reflect on their own learning (2) engage in a specific inquiry (3) question how knowledge is constructed and (4) become self-directed learners.
- *Integration*: students are able to connect ideas in different disciplines, contexts, and realms of life. For instance, students are able to question the statistical uncertainty in a statement, or evaluate data visualizations they see in a news article.
- *Human dimension*: For example, students acquire a better understanding of themselves, learn to contribute to the work of a team, to become an effective leader, or to live with ethical principles. In the context of DSSG, they may find that they enjoy learning with students from diverse backgrounds and they appreciate different perspectives.

- *Caring*: students may become excited about a specific topic, find new interests or values, or become committed to a certain standard. For example, they may become interested in applying data science to enhance the well-being of others, or committed to ethical problem solving and reproducible analysis.

Fink's taxonomy describes learning as a *developmental process*, that involves both intellectual development, as well as changes in behavior, belief, and attitude. Learning "is a developmental process that intersects with other processes in a student's life, and students enter our classrooms not only with skills, knowledge, and abilities, but also with social and emotional experiences that influence what they value, how they perceive themselves and others, and how they will engage in the learning process." [8]

[8] Ambrose Susan A., Bridges Michael W., DiPietro Michele, Lovett Marsha C., and Norman Marie K. *How Learning Works: Seven Research-Based Principles for Smart Teaching*, volume 1 st of *The Jossey-Bass Higher and Adult Education Series*. Jossey-Bass, 2010

**Activity** Connect your big dream for the fellows to Bloom's and Fink's learning objectives. Can you formulate three learning goals for each fellow? As an example, you might want them to be able to "formulate a statistical model and evaluate its appropriateness in a given situation", or "present their results clearly, concisely and confidently".

Now that you have identified learning goals for each fellow, I describe three approaches you can use to help fellows achieve their goals.

## 3.2 *Encouraging peer learning*

Fellows work in groups to tackle a complex problem during DSSG. The *collaborative* nature of this experience provides a great opportunity for them to learn from each other. Below are a few suggestions to support "peer learning". Why do these activities enhance fellows' learning?

### 3.2.1 *Examples of peer learning*

*Peer feedback* During team meeting, you can dedicate some time for fellows to ask each other questions or give feedback. You can ask fellows to review each other's code, and suggest a list of dimensions to examine. For example: Do you notice any errors? Is the code properly documented? How would you improve this piece of code? What did you learn?

*Peer teaching* Identify for each fellow one area that they are strong in and ask them to teach others. [9] You can also ask fellows to explain their approaches to each other.

[9] You may also want to encourage them to reflect on their strength themselves.

*Team work* You can ask fellows to work together on a complex task. For example, each fellow can explore one data set, implement one model, or come up with one way to evaluate a model. [10].

[10] The key to successful team work lies in choosing a task that requires team work, setting up norms, explicitly asking students to share and listen, to make sure their tasks are interdependent while at the same time fellows are responsible for their individual piece of the work (See Chapter L in *The ABCs of How We Learn: 26 Scientifically Proven Approaches, How They Work, and When to Use Them*).

### 3.2.2 *What is peer learning?*

Peer learning is an instruction approach where students work in teams to accomplish a common goal. The team involves five elements: positive interdependence; individual accountability; face to face promotive interaction; appropriate use of collaborative skills; group processing[11]. For instance, students may work on a challenging problem together during a workshop. Before the workshop, they are asked to brainstorm individually; during the workshop, they discuss their solutions, ask clarifying questions and provide feedback to each other. With the guidance of an instructor, they devise a solution to the problem as a group. After the workshop, they might be asked to answer a short quiz to assess their individual understanding of relevant concepts.

Here, I adopt a more general definition of peer learning as any activity where students help each other to learn. The examples we provided before were beneficial to learning because they engage higher level cognitive activities. For example, when students elaborate an approach, they need to retrieve information from memory, connect ideas and explain their reasoning. [12] Peer learning also helps students to build a community of learners. West and Williams (2007) provides one definition of a learning community through four aspects: access (readily accessible to each other), connectedness (sense of belonging, trust and interdependence), shared common purpose towards education and function (such as shared practice and class time) [13]. Zhao and Kuh (2004) observed that students who engaged in a learning community gains more knowledge and competence, are more likely to engage with faculties and are overall more satisfied with their college experience [14]. Research has also shown that students experience more learning gain when engaged with a diverse perspective (See Chapter L in *The ABCs of How We Learn: 26 Scientifically Proven Approaches, How They Work, and When to Use Them*).

[11] Richard M. Felder and Rebecca Brent. Cooperative learning. In Patricia Ann Mabrouk, editor, *Active Learning: Models from the Analytical Sciences (ACS Symposium Series, 970)*, chapter 4. Washington, DC: American Chemical Society, 2007

[12] Chapter T in *The ABCs of How We Learn: 26 Scientifically Proven Approaches, How They Work, and When to Use Them* describes the benefits of explaining and Chapter 4 in *Small teaching: Everyday Lessons From the Science of Teaching* oin]smallteaching2016 describes the benefits of connecting ideas.

[13] Richard E. West and Gregory S. Williams. "I don't think that word means what you think it means": A proposed framework for defining learning communities. *Technol. Res. Dev.*, 65(6):1569–1582, 2017

[14]

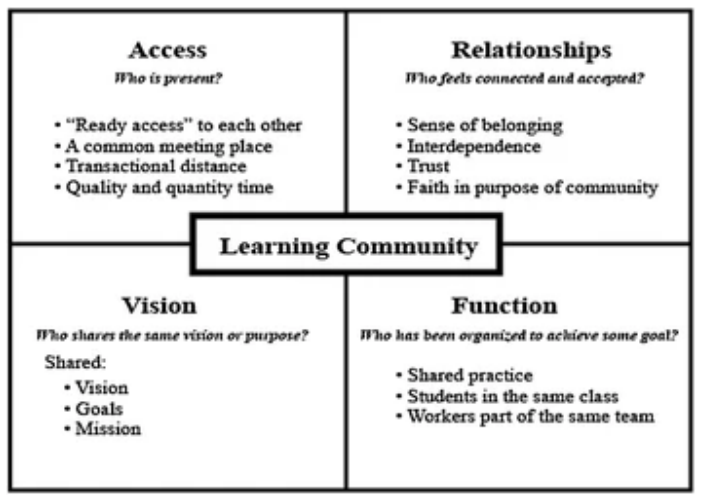

Figure 3.3: Characteristics of a learning community defined in West and Williams (2017).

| |
|---|
| **Activity** Design one activity for your project team that is centered around peer learning. |

## 3.3 *Developing Meta cognition*

Meta-cognition is the process of *reflecting on* and *directing* one's own thinking [15]. Ultimately, we want fellows to become independent data scientists who are able to assess the need of the community partner and carry out a complex project. In this section, I will focus on two aspects of meta-cognition: evaluating one's learning and solving a problem.

[15] National Research Council. *Knowing What Students Know: The Science and Design of Educational Assessment*. The National Academies Press, Washington, DC, 2001

### 3.3.1 *Evaluate one's learning*

Does it every happen to you that you sit down to implement a clustering algorithm that you thought you knew well and all of a sudden unanswered questions appear? Indeed, we often become aware if we know something well when we try to retrieve information. There are many things you can do to help fellows evaluate their learning. Here are some starting ideas and I encourage you to develop more approaches.

- Probe fellow's knowledge by asking them to explain what a particular statistical method is and how it works.
- Ask fellows to document their own work and thinking process.
- Ask fellows to ask each other clarifying questions.

### 3.3.2 *Meta-cognition during problem solving*

Consider the problem of predicting treatment effect on patients, I describe five steps involved in solving this problem and how you can guide fellows to reflect on each step. These steps are based on Chapter 7 in *How learning works: Seven Research-Based Principles for Smart Teaching* and the article *A detailed characterization of the expert problem-solving process in science and engineering: Guidance for teaching and assessment*[16].

[16] Argenta M. Price, Candice J. Kim, Eric W. Burkholder, Amy V. Fritz, and Carl E. Wieman. A detailed characterization of the expert problem-solving process in science and engineering: Guidance for teaching and assessment. *CBE Life Sci. Educ.*, 20(3), 2021

*Framing the problem* Before jumping into finding solutions, fellows need to know precisely what is the problem, what is the goal, and what are some criteria to determine if their solution is successful.

At this stage, we can ask fellows what is the goal of their task and what steps they need to carry out. A successful problem framing might look like "I want to predict a patient's blood pressure during the next twenty-four hours after taking the medication. I choose to look at the next twenty-four hours because that's the time period before they take their next medication. I especially care about how well I am able to capture a hypertensive crisis because of the consequence of missing this event."

*Planning the approach* Next, fellows need to identify appropriate procedures and outline a path towards a solution. This step may involve several iterations, as they may start with one approach, evaluate whether it works, and then experiment with another approach until they decide on a promising method.

At this stage, you can ask them to describe what procedures they plan to use and why they make these selections. You can also demonstrate you thinking process, for example, "I will first simplify the problem — it might be easier to predict if one person would experience hypertensive crisis or not. One way to do that is by using responses of similar individuals. For example, I can use clustering and if I do that, I will first need to compute similarity between each individuals and then I need to choose one specific clustering method. How do I determine similarity between individuals? Perhaps I can use the Euclidean distance, then I can include all of the variables. However, a drawback is that Euclidean distance treats different variables but they may have different importance. I may deal with it by... The different clustering methods I can use are ...".

*Identifying roadblocks and gaps* Next, fellows need to identify gaps in their knowledge and skills. For example, they might need to use a software developed by other researchers, read some literature, or clarify some concepts with the project partner. You can ask fellows to describe what challenges they might encounter and how they can address them. An example of such challenges might be "I need to ask our community partner which features they think are relevant and how to characterize similarity between patients."

*Monitoring progress* After planning the procedure, fellows will implement their plan. At this point, you can encourage them to monitor their own progress. For example, you can ask them to connect what they are currently doing with their plan, what new challenges they encounter, and what additional information they might need. For example, they might notice that several variables are missing, and therefore they need to devise a plan to impute these variables.

*Adjusting one's approach* Finally, fellows need to measure the performance of their solution, interpret their results and reflect on how they can improve or adjust their approach. You can ask them to explain their findings and conclusions. You can ask them to describe how well they think their method is working, and which adjustments they might consider. You can also ask them to connect their progress with the overall project goal. A reflection at this stage may be "the clustering approach gives an accuracy of x%. It's particularly effective at identifying occurrence of hypertensive events because the sensitivity is high. However, this approach does not use the information from individuals who are not similar. The next step is ..."

**Activity** Write down one question to ask fellows during each step in their problem solving. Ask these questions during your team meeting and take notes of a few of their responses. Do you notice any change in how fellows approach a problem during the course of DSSG?

## 3.4 *Incorporating active learning*

Active learning is an instructional method where students are engaged in the learning process, in contrast to the traditional learning approaches where they passively listen to an expert[17]. Students may work in a group, e.g., discussing answers with a partner[18], but they can work individually, e.g., thinking about an answer on their own.

[17] Michael Prince. Does active learning work? a review of the research. *J. Eng. Educ.*, 93(3):223–231, 2004

[18] Scott Freeman, Sarah L. Eddy, Miles McDonough, Michelle K. Smith, Nnadozie Okoroafor, Hannah Jordt, and Mary Pat Wenderoth. Active learning increases student performance in science, engineering, and mathematics. *Proc. Natl. Acad. Sci. U.S.A.*, 111(23): 8410–8415, 2014

Active learning benefits students in many ways. First, it stimulates higher level cognitive process. For example, Freeman et al. (2014) showed that active learning can be used to teach critical thinking skills[19]. In this study, students were asked to take measurements in two different ways and compare the results. Then, they were told to devise a plan to improve measurement quality and carry out the revised measurement. They were then asked to determine whether their data agrees with a model, and if not, whether it is because of data quality or model accuracy, and how would they improve measurement in the first case and how would they carry out new experiment in the second case. By practicing critical thinking, students learned to think like a scientist, who routinely compare results with a model and decide whether to question their measurement or the model. Researchers found that not only is this approach effective, but for some students, the effect persists when they are no longer asked to compare results between measurements.

[19] N. G. Holmes, C. E. Wieman, and D. A. Bonn. Teaching critical thinking. *Proc Natl Acad Sci U S A.*, 112:11199—-11204, 2015

Active learning also helps to increase students' self-efficacy, motivation, and sense of belonging. This is especially true for students from underrepresented minority (URM) groups, who tend to learn better when they have higher self-efficacy[20].

[20] Cissy J. Ballen, Carl Wieman, Shima Salehi, Jeremy B. Searle, and Kelly R. Zamudio. Enhancing diversity in undergraduate science: Self-efficacy drives performance gains with active learning. *CBE Life Sci. Educ.*, 16(4), 2017

In fact, the entire DSSG program is an active learning experience where fellows are engaged in their learning! That's why it's particularly important for you to encourage fellows to work together and discover solutions themselves. In the remainder of this section, I will explore active learning in the narrower setting when you are explaining a new concept.

### 3.4.1 *Examples of active learning*

Below are a few examples of learning activities, and I encourage you to reflect on an effective learning experience you have had in the past.

> **Activity** Think about a learning activity you have participated in, what did you do? Was it effective? How would you improve it?

*Are data i.i.d.?* This was an activity my professor designed for a course on statistical inference. At the start of the lecture, he handed out pieces of paper with sequences of 1s and 0s written on it. Then he asked "Is this sequence of numbers i.i.d. with probability 0.5 being 0 or 1?". We were given 5 minutes to think and then he asked some of us to share our answer to the entire class. After several of us explained our reasoning, he explained that all of the numbers are 0 or 1 with probability one half. However, the numbers are not generated independently because they are more likely to be the same as the preceding number. One way one can discover this is by counting the number of 0s before encountering a 1.

*How much of the Earth is water?* In Section 9.3.3 of *Teaching statistics: a bag of tricks*[21], the authors described the following in-class activity. Students are asked how they would estimate the proportion of earth covered by water. After they mention a few approaches, the instructor would take out an inflated globe and ask students to toss it around, each time, their finger would touch one point on the globe and they are asked to shout out if their finger touched land or water. The instructor would record the answers and then ask students to estimate the proportion of water as well as provide a confidence interval. Finally, the instructor ask students how they would take random samples from a globe.

[21] Andrew Gelman and Deborah Nolan. *Teaching Statistics: A Bag of Tricks*. Oxford University Press, 2 nd edition, 2017

*Alien attack* This example is from the article *Why we are teaching science wrong, and how to make it right* [22]. During a neuroscience class, the instructor asks students to answer the following question and explain their reasoning: You're innocently walking down the street when aliens zap away the sensory neurons in your legs. What happens? (a) Your walking movements show no significant change. (b) You can no longer walk. (c) You can walk, but the pace changes. (d) You can walk, but clumsily.

[22] M. Mitchell Waldrop. Why we are teaching science wrong, and how to make it right. *Nature*, 523(7560):272–274, 2015

### 3.4.2 *How to incorporate active learning?*

As you can see from the examples, you can incorporate active learning through designing an activity in which fellows can participate, or through asking a thought-provoking question.

In the article *Learning by Doing*, the authors offer several simple ways to integrate active learning in your teaching[23] (the following is an excerpt from the article):

[23] Richard M. Felder and Rebecca Brent. Learning by doing. *Chem Eng Educ*, 37 (4):282–283, 2003

1. Give the students something to do (*answer a question, sketch a flow chart or diagram or plot, outline a problem solution, solve all or part of a problem, carry out some formula derivation, predict a system, interpret an observation or an experiment result, critique a design, troubleshoot, brainstorm, come up with a question, ...*).

2. Tell them to work individually, in pairs, or in groups of three or four; tell them how long they'll have (anywhere from 10 seconds to two minutes); and turn them loose.

3. Stop them after the alloted time, call on a few individuals for responses, ask for additional volunteered responses, provide your own response if necessary, and continue teaching.

The article also describes the classical think-aloud pair problem solving (TAPPS) approach: students are asked to work on a problem in pairs, one of them explaining a solution and the other student ask questions. After sometime, you can ask students to share their answers, and you can provide your solution if necessary. Then, you can ask students to switch roles.

Now that you have seen a few examples of active learning, I invite you to design one activity for your tutorial.

**Activity** Design one activity for your tutorial session. What's the goal of this activity? Write down a prompt, what should fellows do during the activity?

# 4 Acknowledgements

Materials in this handout are inspired by several courses I have taken at Stanford: *Science and Engineering Course Design*, taught by Sheri Sheppard and Diane Lam; *Inclusive Mentoring in Data Science*, taught by Chiara Sabatti and guest lecturers including Mercella Anthony and Joseph Brown. These two courses have taught me much about teaching and mentoring.

I have participated in Stanford Data Science for Social Good for three times, first as a fellow and then as a technical mentor. I am grateful to my mentors during the inaugural DSSG, Ben Stenhaug, Michael Sklar and Emily Flynn, who not only helped me develop R programming skills, but also encourage me to pursue teaching as part of my future career. I am grateful to my fellow mentors, Shilaan Alzahawi, Lijing Wang, Min Woo Sun, and our head technical mentors (Faidra Monachou, Armin Thomas, Kiran Shiragur, Yan Min), for their support and inspiration throughout the program. Finally, I deeply appreciate our faculty advisors, Chiara Sabatti and Balasubramanian Narasimhan, for being wonderful role models, for their continuous encouragement and their generous advise, and for their efforts that make this program possible.